\documentclass{aa}

\newcommand{\Omegab}{\Omega_\text{b}}
\newcommand{\kmskpc}{kms$^{-1}$kpc$^{-1}$}
\newcommand{\kmsdeg}{kms$^{-1}$deg$^{-1}$}
\newcommand{\kms}{kms$^{-1}$}

\usepackage[normalem]{ulem}
\usepackage{graphicx}
\usepackage{txfonts}
\usepackage{algorithm,algorithmic}
\usepackage{xcolor}
\usepackage{bm}
\defcitealias{Ramos_2018}{R18}

\begin{document}

   \title{From ridges to manifolds: 3D characterization of the moving groups in the Milky Way disc}

   \author{M. Bernet \inst{1,2,3}
            \and P. Ramos \inst{1,2,3,4}
            \and T. Antoja \inst{1,2,3}
            \and B. Famaey \inst{4}
            \and G. Monari \inst{4}
            \and \\ H. Al Kazwini \inst{4}
            \and M. Romero-G\'omez \inst{1,2,3}
          }
    \institute{Departament de Física Qu\`antica i Astrof\'isica (FQA), Universitat de Barcelona (UB),  C Mart\'i i Franqu\`es, 1, 08028 Barcelona, Spain
           \email{mbernet@fqa.ub.edu}
    \and{Institut de Ci\`encies del Cosmos (ICCUB), Universitat de Barcelona (UB), C Mart\'i i Franqu\`es, 1, 08028 Barcelona, Spain}
    \and{Institut d'Estudis Espacials de Catalunya (IEEC), C Gran Capit\`a, 2-4, 08034 Barcelona, Spain} 
    \and{Universit{\'e} de Strasbourg, CNRS, Observatoire astronomique de Strasbourg, 11 rue de l’Universit{\'e}, 67000 Strasbourg, France}\\
    }

   \date{Received YYY; accepted XXX}

 
  \abstract
   {The details of the effect of the bar and spiral arms on the disc dynamics of the Milky Way are still unknown. The stellar velocity distribution in the Solar Neighbourhood displays kinematic substructures, which are possibly  signatures of these processes and of previous accretion events.
   With the Gaia mission, more detail of these signatures, such as ridges in the $V_{\phi}$-$R$ plane and thin arches in the $V_\phi$-$V_R$ plane, has been revealed. The positions of these kinematic substructures --moving groups-- can be thought of as continuous manifolds in the 6D phase space, and the ridges and arches as specific projections of these manifolds.}
   {We aim to detect and characterize the moving groups along the Milky Way disc, sampling the Galactocentric radial and azimuthal velocities of the manifolds through the 3 dimensions of the disc: radial, azimuthal and vertical.}
   {We develop and apply a novel methodology to perform a blind search for substructure in the Gaia EDR3 6D data, which consists in the execution of the Wavelet Transform in independent small volumes of the Milky Way disc, and the grouping of these local solutions into global structures with a method based on the Breadth-first search algorithm from Graph Theory. We apply the same methodology to simulations of barred galaxies to validate the method and to compare with the data.
   }
   {We reveal the skeleton of the velocity distribution, uncovering projections that were not possible before.
   We sample nine main moving groups along a large region of the disc in configuration space, covering up to $6$\,kpc, $60$\,deg, and $2$\,kpc in the radial, azimuthal, and vertical directions respectively, extending significantly the range of previous analysis.
   In the radial direction, we find that the groups deviate from the lines of constant angular momentum that one would naively expect from an epicyclic approximation analysis of the first order effects of resonances. In fact, we reveal that the spatial evolution of the moving groups is complex and that the configuration of moving groups in the Solar Neighbourhood is not maintained along the disc.
   We also find that the azimuthal velocity of the moving groups that are mostly detected in the inner parts of the disc (\emph{Acturus}, \emph{Bobylev}, and \emph{Hercules}) is non-axisymmetric. For \emph{Hercules}, we measure an azimuthal gradient of $-0.50$\,\kmsdeg\,at $R = 8$\,kpc.
   We detect a vertical asymmetry in the azimuthal velocity for the \emph{Coma Berenices} moving group which is not expected for structures originating from a resonance of the bar, supporting the previous hypothesis of the incomplete vertical phase-mixing of the group. In our simulations, we extract substructures corresponding to the Outer Linbdlad Resonance and the 1:1 resonances and observe the same deviation from constant angular momentum lines and the non-axisymmetry of the azimuthal velocities of the moving groups in the inner part of the disc.}
   {This data-driven characterization is a starting point for a holistic understanding of the moving groups. It also allows for a quantitative comparison with models, providing a key tool to comprehend the dynamics of the Milky Way.}
   
   \keywords{Galaxy: disc --
            Galaxy: kinematics and dynamics --
            Galaxy: structure -- 
            Galaxy: evolution --
            Methods: data analysis
            }

   \maketitle


%

\section{Introduction}


The stellar velocity distribution in the solar neighbourhood (SN) has been for a long time a key element in our understanding of the structure of the Milky Way (MW) \citep{dehnen1998,skuljan1999,famaey2005kmgiants,antoja2008origin}. Historically, several overdensities in this velocity distribution have been identified and discussed (Pleiades, Hyades, Sirius). These moving groups, as they are usually referred to, can be related to the orbital resonances of the bar and spiral arms of the Galaxy \citep{kalnajs1991pattern,dehnen2000effect,antoja2011understanding,fragkoudi2019ridges,monari2019signatures} and/or attributed to ongoing phase mixing related to external perturbations \citep{minchev2009milky,gomez2012signatures,antoja2018dynamically,Ramos_2018,hunt2018transient,khanna2019galah,laporte2019footprints,laporteages}.

The latest releases of the \emph{Gaia} mission \citep{DR2,EDR3} have provided a full 6D phase-space catalogue of 7.2 million stars, increasing the size and precision of any previous survey by several orders of magnitude. This has been a game-changer in many fields of astrophysics. In the SN, the new high resolution velocity distribution has revealed a complex substructure, with several thin arches never observed before \citep{katz2018dr2VR}. When extending the study to the entire disc, large ridges appeared in the $R$-$V_\phi$ (respectively, Galactocentric radius and azimuthal velocity) diagram covering several kiloparsecs \citep{antoja2018dynamically,kawata2018radial,fragkoudi2019ridges}.

Orbits in a barred potential can be trapped into resonances \citep{weinberg1994resonance}. 
\citet{dehnen2000effect} showed that in a short/fast bar scenario (i.e. $\Omegab = 50$\,\kmskpc) the transition between two types of non-axisymmetric orbital families across the bar's Outer Lindblad resonance (OLR) can explain the bi-modality formed by \emph{Hercules} and the rest of the velocity distribution in the solar neighbourhood if the Sun is placed just outside the OLR of the bar ($R_{\rm{OLR}} \approx 7.2$\,kpc). This scenario was consistent with the gas dynamics measurements of the inner MW at the time. Later on, studies of star counts and kinematics of the inner MW suggested that the the bar might be longer and slower than previously thought \citep{portail2017}. In this case, the OLR would be placed further out ($R_{\rm{OLR}} \approx 10.5$\,kpc, maybe matching other groups such as the \emph{Arch/Hat} instead of \emph{Hercules}) and co-rotation (CR) would be closer to the SN ($R_{\rm{CR}}\approx6$\,kpc). \citet{perez2017revisiting} and \cite{monari2019signatures} then explained \emph{Hercules} as the overdensity formed by the orbits trapped at the CR, librating around the Lagrangian points of a long/slow bar. This moving group created by CR seems to be less pronounced than the one produced by the OLR \citep{binney2018orbits, hunt2018hercules}. However, \citet{hunt2018transient} showed that the addition of spiral structure in combination with the CR might create a strong distinct Hercules, and \citet{chiba2021decelerating} showed that a decelerating Galactic bar could enhance the occupation on resonances, being able to reproduce Hercules thorugh the CR resonance. This shows that the value of the pattern speed ($\Omegab$) of the MW bar as well as the exact link between substructures and resonances are still a matter of debate, and more observables are needed to obtain a final answer.

In this direction, \citeauthor{Ramos_2018} (\citeyear{Ramos_2018}, hereafter \citetalias{Ramos_2018}), used the wavelet transform \citep[WT,][]{starck2002astronomical,chereul1999distribution} to detect and characterize the kinematics of the moving groups along the disc. They matched the spatial evolution of the groups with the ridges in the $R$-$V_\phi$ plane. They claimed that some of the arches follow lines of constant energy at a given volume --which could be related to phase mixing processes \citep{minchev2009milky,gomez2012signatures}-- and others follow lines of common angular momentum in the radial direction, as expected approximately in the case of resonant kinematic substructures \citep[e.g.,][]{quillen2018spiral}. They also claimed that the observed changes in the azimuthal direction for the Hercules moving group are consistent with being produced by the OLR of a short/fast bar \citep{dehnen2000effect,fux2001order,antoja2014constraints}. The long/slow paradigm is relatively recent and there have been few analyses on the azimuthal variations of a substructure caused by CR. \citet{monari2019hercules} found that the Hercules angular momentum changes significantly with azimuth as they predicted analytically for the co-rotation resonance of an old long/slow bar. They showed that the only way to obtain a similar change in azimuth for an OLR origin of Hercules would be if orbits are still far from phase-mixed in the bar potential \citep[bar perturbation younger than $2$ Gyr; see also][]{Trick_2021}.

The link between the moving groups across the neighbourhoods in \citetalias{Ramos_2018} was made visually, using a scatter plot of two variables and a third one as colour. Therefore, the analysis of the moving groups link was restricted to three variables. Since $V_R$ and $V_\phi$ are compulsory to select the moving groups, this limitation restricted the analysis to one dimension in space (either radial or azimuthal). The vertical direction was not explored.

The correlation between the position of the moving groups (overdensities in $V_R - V_\phi$) and the ridges (overdensities in $R - V_\phi$) indicate that both are projections of the same substructure in the 6D phase-space onto different planes. The positions of these kinematic substructures --moving groups-- can be described as continuous manifolds in the 6D phase space, and the ridges and arches as specific projections of these manifolds. Our goal in this article is to extend the idea introduced in \citetalias{Ramos_2018} by automatising the n-dimensional link of the moving groups to avoid the limitation of projecting the data. With this, we intend to move from a ridge-moving group paradigm to a manifold paradigm, were we sample the position of these manifolds in the $(R,\phi,Z,V_R,V_\phi)$ space for each moving group.

We present a novel methodology to detect these manifolds in a dataset. It is based on the execution of the WT in independent small volumes, and the relation of these local solutions in global substructures with an algorithm based on the Breadth-first search (BFS) algorithm from Graph Theory. With this methodology, we process the Gaia EDR3 6D data and detect the positions of the groups across the MW disc. We also sample the manifolds of two test particle simulations with a fast ($\Omegab = 50$\,\kmskpc) and a slow ($\Omegab = 30$\,\kmskpc) bar, both to test our methodology and to compare it to the data.

The Gaia DR3 \citep{gaiaDR3} catalogue includes a larger and updated sample of radial velocities \citep[33 M of stars,][]{drimmel2022dr3_6d}, which covers a larger region of the MW disc and increases the resolution (number of stars and precision) of Gaia EDR3. This provides finer observables to untangle the different contributions in the complex dynamics of the Galaxy. To exploit these data in its totality, new strategies must be developed \citep[e.g.][]{contardo2022emptiness} to avoid the current limitations in the analysis, which the present article contributes to.

This paper is organized as follows. In Section \ref{sect:data}, we describe the observational data that we used. In Section \ref{sect:method}, we introduce the methodology we developed. In Section \ref{sect:results}, we show the results of the application of the method to Gaia EDR3 data. In Section \ref{sect:simulation}, we present and analyse the simulations. In Section \ref{sect:discussion}, we compare the results from the data and the simulations, and with previous results in the literature. Finally, in Section \ref{sect:conclusions} we list the main conclusions of this work.

%
%

\section{Data and sample preprocessing}\label{sect:data}

The Early release of Gaia DR3 consists of an updated and enlarged source list, with improved astrometry and photometry. Besides proper motions, about 7.2 million stars have radial velocity (RV) measurements in the Gaia DR2, most of which are transferred to EDR3 \citep{seabroke2021radial,torra2021gaia}. For this section, we use a subset of these stars with photo-geometric distances from \citet{bailer2021estimating}, derived
from a probabilistic approach including colour and apparent magnitude information. 

Distance is a critical parameter in the computation of the motion and position of the star in the 6D study of the MW, and a major source of uncertainty. This is why, in order to improve the quality of the sample, we additionally apply a cut in relative parallax error:
\begin{equation}
    \frac{\varpi}{\sigma_{\varpi}}> 5.
\end{equation}
The resulting sample contains $6\,059\,648$ sources.

In the Appendix~\ref{sect:distance_bias}, we study the errors that an overestimation or an underestimation of the distance could produce in our results. We determine that, in general, this error would be below $2$\,\kms\,for a distance bias of $\pm10\%$, and it would not affect the overall trend of the groups.

We use a Cylindrical Galactocentric coordinate system, fixing the reference at the Galactic Centre (GC) with the radial direction ($R$) pointing outwards from it, the azimuthal ($\phi$) negative in the direction of rotation, and the vertical component ($Z$) positive towards the North Galactic Pole. To transform Gaia observables to positions and velocities in this reference frame, we take the Sun to be at $R_\odot = 8.178$\,kpc \citep{abuter2019geometric}, $\phi_\odot = 0º$ and $Z_\odot = 0.0208$\,kpc \citep{bennett2019vertical}. For the solar motion, we use $U_\odot = 11.1, v_{circ} + V_\odot = 248.5, W_\odot = 7.25$\,km\,s$^{-1}$ \citep{schonrich2010,reid2020}.

%
%

\section{Method}\label{sect:method}

The data described in the previous section contains the 6D variables of position $(R,\phi,Z)$ and velocity $(V_R,V_\phi,V_Z)$ of the stars. Inside small volumes --cuts in $(R,\phi,Z)$--, the moving groups appear as well defined overdensities in the velocity distribution $V_R$-$V_\phi$ \citepalias{Ramos_2018}, which are easy to detect. However, we know that at large spatial scales the position of the overdensities in the velocity space changes (ridges in $R$-$V_\phi$). Therefore, if we use larger volumes to construct the velocity distribution the overdensities will blur and become undetectable.

In this section, we present the novel method that we developed to extract these large kinematic substructures from a dataset. It is divided into two steps; the execution of the WT in independent small volumes of the MW disc, and the relation of these local solutions in global substructures, with an algorithm based on the Breadth-first search (BFS) algorithm from Graph Theory \citep[descrived in Section \ref{sect:BFS}]{moore1959bfs}.

\subsection{Local Wavelet Transform}\label{sect:WT}

We partition the data in a dense grid of small volumes (from now, \emph{pixels}) in the spatial coordinates. We construct it as follows:
\begin{itemize}
    \item Radial direction $(R_i)$: $[5,14]$ kpc in steps of $0.04$ kpc,\\ $R_{bin} = \pm0.24$ kpc around each centre
    \item Azimuthal direction $(\phi_j)$: $[-34,34]$ deg in steps of $0.8$ deg,\\ $\phi_{bin} = \pm2.4$ deg around each centre
    \item Vertical direction $(Z_k)$: $[-1,1]$ kpc in steps of $0.08$ kpc,\\$Z_{bin} = \pm0.24$ kpc around each centre
\end{itemize}
which produce a dense grid of $2\,700\,000$ pixels, with a maximum volume overlap between consecutive pixels of $83.3\%$. This bin size is bigger than the one used in \citetalias{Ramos_2018}. When reaching regions far from the Sun, the statistical significance of the moving groups decreases and a bigger bin is the only way to obtain a robust determination of the group velocity. The drawback for this enlarged bin is that any inhomogeneity of the sample within a volume (due to the local extinction, the selection function, a change in the sampled populations, etc.) can bias the mean position of the stars and therefore its kinematics. We tested this effect using a smaller bin, and we estimate the impact to be below $2$\,\kms\,for the used bin size.

\begin{figure*}
\centering
\includegraphics[width=0.99\textwidth]{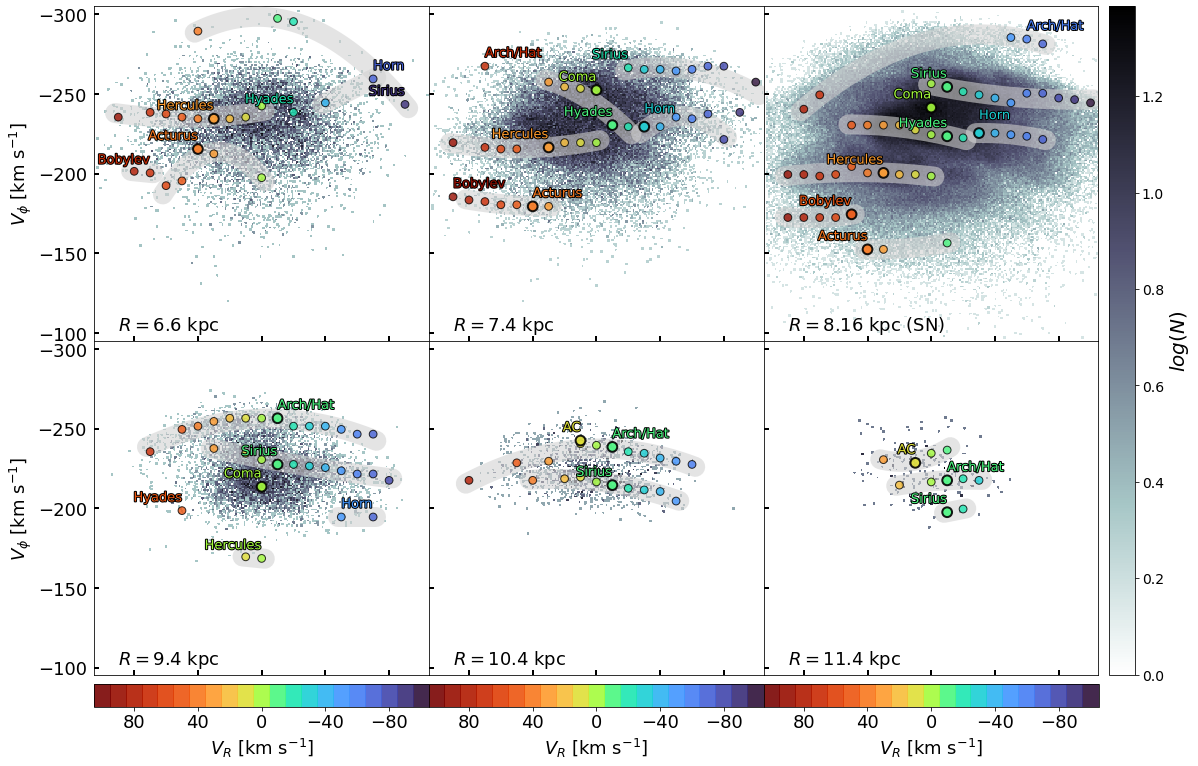}
\caption{Moving group detection in different neighbourhoods along the radial direction. For each moving group, we include a parabolic fitting of the substructures associated with each group in a thick grey line. Each moving group contains several structures, corresponding to different $V_R$ bins. The largest structure in each group is used as its representative (dots with larger black contours and the moving group name on top). We can see two examples of bimodalities, which serve as evidences of the complex evolution of the arch morphology (\emph{Sirius} at $R=8.16$\,kpc and \emph{Arch/Hat} at $R=10.4$\,kpc).}
\label{fig:SN_WT}
\end{figure*}

For each pixel, we construct the velocity distribution $(V_R,V_\phi)$ diagram of the stars in it as a 2D histogram with bins of $1$\,\kms\, (see background histogram in Fig.~\ref{fig:SN_WT}). \citetalias{Ramos_2018} showed that the overdensities form thin arches elongated around large ranges of $V_R$, with a small variation in $V_\phi$. The use of 2D peak detection algorithms \citepalias[as the one in][]{Ramos_2018} is sub-optimal for arch-like structure detection. When analysing regions with few observations, the search for peaks is translated into a very noisy determination of $V_R$, and uncontrollable correlations between $V_R$ and $V_\phi$ (movement along the arch). 

To avoid this, we slice each $V_R - V_\phi$ diagram in vertical columns (bins in $V_R$), and run a 1D WT in the $V_\phi$ histogram of each column:
\begin{itemize}
    \item Radial velocity ($V_R$): $-100$ -- $100$\,\kms\, in steps of $10$\,\kms, \\ $V_{R\,bin} = \pm15$\,\kms\, around each centre
\end{itemize}

Since we are detecting each part of the arch separately, we avoid the movement along the arch of the overdensities, breaking the degeneracy between $V_R$ and $V_\phi$ in the detection. To detect the peaks, we use the algorithm developed in \citet{Du2006wavelet1d} implemented in \texttt{scipy} \citep{scipy2020} as \texttt{find\_peaks\_cwt}. This method performs the 1D WT in a range of length scales. A peak is then selected if it is present in enough scales consecutively. In our execution, we use a range of scales of $[5,10]$\,\kms, with steps of $1$\,\kms. We keep the peak if it is present in more than two scales consecutively. With this configuration of scales, we loose the thin resolution that we could extract in regions with a large number of sources, but we gain robustness in the detection of the large structures in poorly sampled regions. Since the scope of this work is the large-scale behaviour of the groups, we consider this approach to be better.

At the end of the execution, the peak $p$ inherits the spatial position from the pixel, $V_R$ from the position of the radial velocity bin, and $V_\phi$ from the result of the WT detection. Therefore, a peak has the coordinates
\begin{equation}\label{eqn:coordinates}
p = (R,\phi,Z,V_R,V_\phi)_p.
\end{equation}

\subsection{Breadth-first search (BFS) resolution with online interpolator}\label{sect:BFS}

We have defined the pixels to have a large overlap among them (two adjacent pixels will share $83.3\%$ of their volume). Therefore, a given substructure in consecutive pixels will have an almost identical shape.

For a pair of peaks from consecutive pixels, we consider them to be adjacent if they are in the same $V_R$ bin and their distance in $V_\phi$ is smaller than $4$\,\kms. In \citetalias{Ramos_2018}, the maximum slope found in a moving group is $33$\,\kmskpc. In our grid the step is $0.04$\,kpc, which translates in a maximum change of $\approx 1$\,\kms\,between adjacent pixels. This $4$\,\kms\,limit in the adjacency is a compromise between including very steep groups (about 4 times the one detected in \citetalias{Ramos_2018}) and reducing the number of adjacencies, which will determine the computational cost of the next step. 

In some occasions, especially in poorly sampled regions, a peak from one pixel can be exactly in the middle of two peaks in the other pixel. In these cases, we do not consider any of them adjacent. With this consideration, two adjacent peaks are always strong candidates to belong to the same substructure.

This adjacency information constructs an enormous net of linked peaks. Ideally, substructures will be isolated subsets of peaks in this net. These are groups of peaks with no adjacencies to any peak outside their group.

In Graph Theory, these nets of linked points are called Graphs and the isolated groups are the connected components of a Graph. A very common algorithm to extract these connected components is the Breadth-first search (BFS) algorithm, which proceeds as follows:
\begin{enumerate}
    \item Add (enqueue) the initial peak $p$ to the queue\footnote{A queue is a data structure similar to an array with limited access to the positions. One end is always used to insert data (enqueue) and the other is used to remove data (dequeue). Queue follows First-In-First-Out methodology, i.e., the data item stored first will be accessed first.} Q.
    \item Select (dequeue) the top peak $p_{top}$ of the queue Q.
    \item Visit all the peaks $p_{adj}$ adjacent to $p_{top}$. For each adjacent peak, if we have already visited it, ignore the peak. If it is the first time we see the peak, enqueue it.
    \item If there are still peaks in the queue, return to step 2.
    \item If the queue is empty, our connected component is the list of visited peaks.
\end{enumerate}

Given an initial peak $p_{0}$, Algorithm \ref{algo:bfs_simple} (see below) returns the entire substructure to whom it belongs. By repeating the process for all the non-matched peaks we can extract all the substructures.

\begin{algorithm}
\caption{-- Breadth-first search}
\label{algo:bfs_simple}
\begin{algorithmic}[1]
\STATE{queue $Q$}
\STATE{list $V$}
\STATE{add $p_{0}$ to $V$ and enqueue in $Q$}
\WHILE{$Q$ not empty}
    \STATE{$p_{it} \leftarrow$ dequeue $Q$ {\it (remove and assign)}}
    \FORALL{$p_{adj}$ adjacent to $p_{it}$}
        \IF{$p_{adj}$ not in $V$}
            \STATE{add $p_{adj}$ to $V$, enqueue in $Q$}
        \ENDIF
    \ENDFOR
\ENDWHILE
\RETURN $V$
\end{algorithmic}
\end{algorithm}

This solution would be enough in an ideal case, but in practice undersampling and Poisson noise especially in regions far from the Sun produce confusion and jumps between structures that a straightforward BFS implementation can not filter out.

In order to avoid these jumps between structures, we include an extra step in the algorithm. While the BFS is running, the peaks already matched give us information about the structure. Therefore, in order to accept a new peak, we will require it to be consistent with the current structure.

Let us suppose we have a group $V$ of already visited peaks (the current substructure we are extracting). To see if a peak $p$ is consistent with this substructure we will select all the peaks in $V$ in a small subset $S\subset V$ around $p$. With this local sample we can compute a linear fit 
\begin{eqnarray}\label{eqn:linear_approx}
    V_\phi \approx a_0 + a_1  R + a_2 \phi + a_3 Z
\end{eqnarray}
of the subset $S$ around $p$ and predict the expected $V_\phi$ of the substructure in a given position. This works under the assumption that the manifold that follows the substructure is derivable and we can compute its first order approximation locally.

This prediction is already absorbing the offset in the structure position produced by its slope in a certain direction. Therefore, the criteria in the acceptance of a new peak should be more strict than the one in the first adjacency step. Our limit in the resolution is the $1$\,\kms\,bin in the $V_\phi$ histogram, and we include an extra tolerance of $0.5$\,\kms. If the distance between the peak azimuthal velocity $V_{\phi,p}$ (Eq. \ref{eqn:coordinates}) and the prediction is smaller than $1.5$\,\kms, we consider the peak to be consistent with the structure. We encapsulate this in the \textbf{is\_consistent\_with} function (Algorithm \ref{algo:is_consistent_with}). We provide a summary of the final algorithm in pseudocode (Algorithm \ref{algo:bfs_simple_interpolator}).

\begin{algorithm}
\caption{-- is\_consistent\_with($p_{adj},V$)}
\label{algo:is_consistent_with}
\begin{algorithmic}[1]
\STATE{$S = V\big(|R_{p_{adj}}-R_V|<R_{fit} \quad \&$\\ \quad \quad \quad $|\phi_{p_{adj}}-\phi_V|<\phi_{fit} \quad \&$ \\ \quad \quad \quad $|Z_{p_{adj}}-Z_V|<Z_{fit}\big)$}
\STATE{$f(R,\phi,Z)=V_\phi \leftarrow linear\_fit(S|_{p_{adj}})$}
\IF{$|f((R,\phi,Z)_{p_{adj}}) - V_{\phi,p_{adj}}|<d$}
    \RETURN{True}
\ELSE
    \RETURN{False}
\ENDIF
\STATE{$R_{fit} = 1$\,kpc, $\phi_{fit} = 4$\,deg, $Z_{fit} = 0.2$\,kpc, and $d = 1.5$\,\kms.}
\end{algorithmic}
\end{algorithm}

\begin{algorithm}
\caption{-- Breadth-first search with online interpolator}
\label{algo:bfs_simple_interpolator}
\begin{algorithmic}[1]
\STATE{queue $Q$}
\STATE{list $V$}
\STATE{add $p_{0}$ to $V$ and enqueue in $Q$}
\WHILE{$Q$ not empty}
    \STATE{$p_{it} \leftarrow$ dequeue $Q$ {\it (remove and assign)}}
    \FORALL{$p_{adj}$ adjacent to $p_{it}$}
        \IF{$p_{adj}$ not in $V$ and \textbf{is\_consistent\_with}($p_{adj},V$)}
            \STATE{add $p_{adj}$ to $V$, enqueue in $Q$}
        \ENDIF
    \ENDFOR
\ENDWHILE
\RETURN $V$
\end{algorithmic}
\end{algorithm}

%
%

\section{Results}\label{sect:results}

Within the 3D grid the method extracted hundreds of structures, covering $2.5$ to $6$\,kpc in $R$ and $30$ to $60$\,deg in $\phi$. In \citetalias{Ramos_2018} the arches in the solar neighbourhood and the radial direction were carefully characterized, and matched to the groups previously studied in the literature. We rely on this matching to associate the results of our methodology with the different groups and arches.

We end up with a sample of $99$ structures, each one associated with one of the nine main moving groups: \emph{Acturus}, \emph{Bobylev}, \emph{Hercules}, \emph{Horn}, \emph{Hyades}, \emph{Sirius}, \emph{Coma Berenices}, \emph{Arch/Hat} \citepalias[][references therein]{Ramos_2018}, and \emph{AC} \citep[see Anti-Centre newridge 1 in][]{antoja2021anticentre}. Each structure traces the position of the moving group along the space at a given $V_R$. Therefore, each group of structures traces the manifold of the position of the moving groups in the $(R,\phi,Z,V_R,V_\phi)$ space. For each moving group, we select its largest structure as its representative. These representatives will be used in the following sections to study the behaviour of the groups in $R-\phi$ and $R-Z$ projections.

In Fig.~\ref{fig:SN_WT}, we show the selected groups in several neighbourhoods in the radial direction. In each arch we highlight its representative with a larger black border. As explained in Section \ref{sect:WT}, our goal in this work is to be able to perform this analysis in a large extent of the disc, so we tuned the detection parameters to obtain a robust detection of the main structures in noisy regions (see $R>10$\,kpc in Fig.~\ref{fig:SN_WT}), and this required to use larger spatial bins. Because of this, the thin arches observed in the Gaia velocity distribution in the SN are slightly washed out.

The methodology links the structures in a given $V_R$, but does not provide the link of the arches in $V_R$-$V_\phi$. The complex nature of the arches formed by the moving groups at different positions in the disc and the high level of noise result in a sub-optimal global link of the arches in the velocity distribution.
Therefore, we only provide a tentative manual linking of the structures, based on the study of \citetalias{Ramos_2018}. In the rest of the paper we use this arch link as a qualitative tool in the analysis, but the main conclusions are based on the properties of the individual parts of the arches, which are determined by the described methodology.

This linking provides interesting results, different than in previous studies. For instance, the evident link of the \emph{Arch/Hat} at $R=9.4$\,kpc creates an asymmetric arch shape for this structure at the SN. This could be an artifact of the detection of \emph{Arch/Hat} at $V_R>60$\,\kms\, or a physical evidence of an unknown behaviour related to its origin. Notice that tagging the groups in the the SN and assuming that they keep united across the disc is a clear oversimplification. In the data, we have seen that \emph{Sirius} is formed by two arches at SN but these arches merge in a single one at the outer parts of the disc (pannels $R=8.16,9.4$\,kpc in Fig.~\ref{fig:SN_WT}). The same happens for \emph{Arch/Hat} at $R=10.4$\,kpc. In the simulation (Sect. \ref{sect:simulation}), we observe the same behaviour for overdensities related to the Outer Linblad Resonance.
So far, this simplification is useful for the discussion and comparison to the state of the art. Moreover, the lack of data far away from the SN does not allow a robust arch characterization. In future releases, an automatized arch detection will be needed to disentangle the complex orbit distribution.

%
%

\begin{figure*}
\centering
\includegraphics[width=1\textwidth]{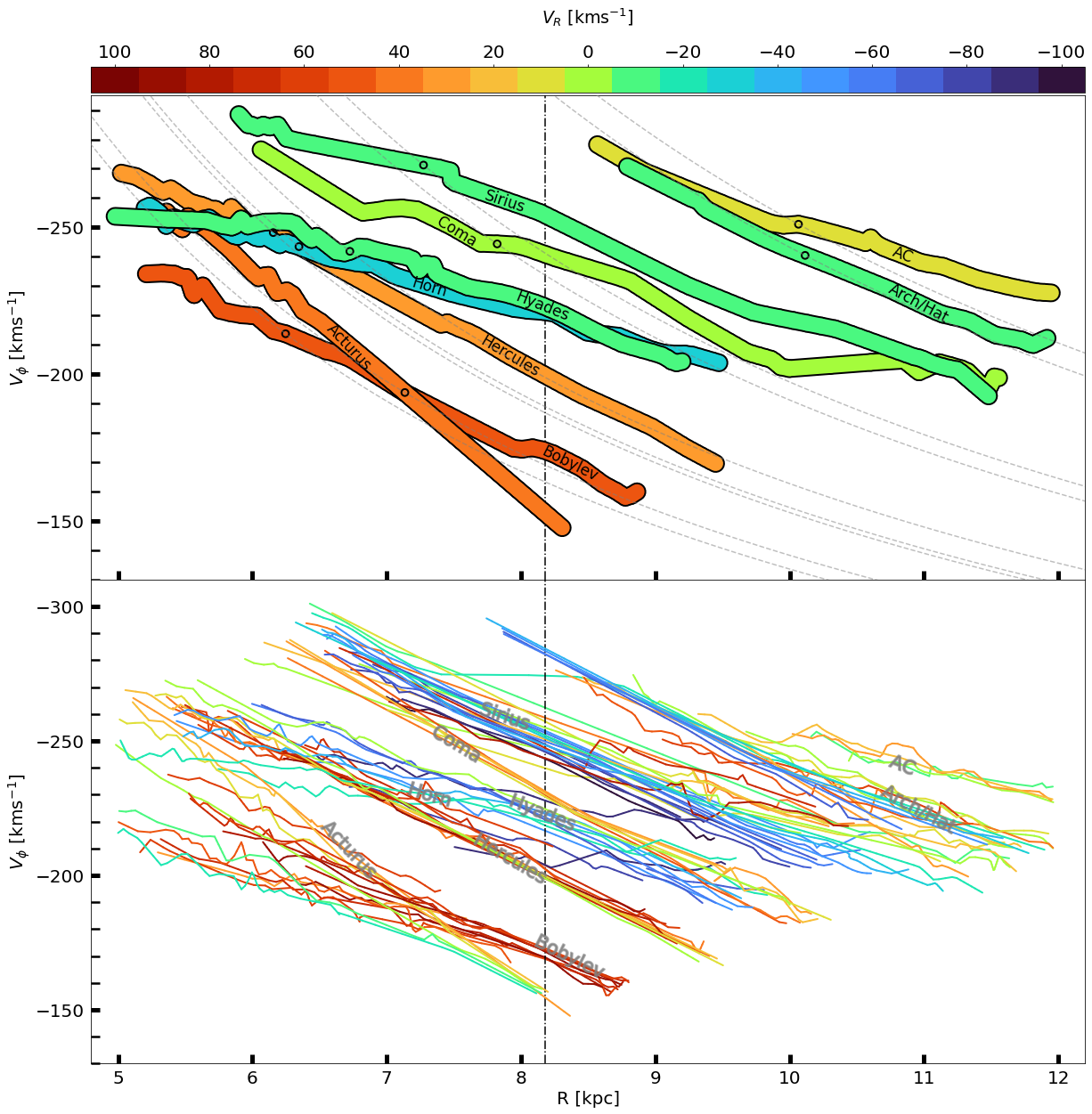}
\caption{Azimuthal velocity of the kinematic substructures in the radial direction, $\phi=0$º, $Z=0$\,kpc, as a function of the radius, and coloured by their radial velocity. The dashed grey lines correspond to constant angular momentum lines which provide a better fit for each structure. Top: Structures corresponding to the main peak of a moving group, tagged with the name in literature. Bottom: Secondary peaks of the moving groups. The usual way to observe this projection is using the number of stars or the mean $V_R$ in each bin \citep[see Fig.~1 in][]{fragkoudi2019ridges}. Using our methodology we can observe the skeleton of the distribution and its complexity. For instance, the slope of the moving groups at different $V_R$ can be very different and the groups cross each other or share close regions in the diagram. In addition, the extension of the range of exploration shows that the moving group deviate from the constant vertical angular momentum predicted for small epicyclic amplitudes.}
\label{fig:r_vphi_groups}
\end{figure*}

\subsection{Radial direction}\label{sect:radial_direction}

The first evidence of large-scale substructure in the dynamics of the disc was the presence of ridges in the $R$-$V_\phi$ plane, directly related to the moving groups observed in the SN. Therefore, a first exercise we can do with the manifolds is to extract their subsets in the radial direction (i.e. $\phi=0$\,deg, $Z=0$\,kpc) and plot them in $R$-$V_\phi$ plane, colored by $V_R$ (Fig.~\ref{fig:r_vphi_groups}). In the top panel, we show the representative groups, tagged by their literature name. In the bottom panel, we show the rest of structures as beams of lines that define the morphology of the corresponding moving groups.
As expected, when tracing the different moving groups along $R$ we observe diagonal lines in $R$-$V_\phi$, matching the already known ridges.

Comparing with \citetalias{Ramos_2018}, we detect the same moving groups but we manage to extend their detection several kiloparsecs. The results in the inner and outer part of the disc ($R<6.5$\,kpc and $R>10$\,kpc) are noisy due to Poisson noise. The Gaia DR3 release will improve the detection of groups in these regions, but even if we exclude this part the groups extend far beyond the range seen in other studies \citepalias[see Fig.~6 in][]{Ramos_2018}.
This extension of the structures is due to both a major improvement of the methodology and the use of the updated astrometric Gaia EDR3 data. The lower error in proper motion and parallax increases the concentration of the moving groups in the undersampled regions.

In Fig.~\ref{fig:r_vphi_groups}, we can observe how the slope of the lines in radius is not constant across the different groups. In the top plot, groups like \emph{Acturus}, \emph{Hercules}, and \emph{Arch/Hat} present slopes significantly steeper than \emph{Sirius} and \emph{AC}. However, this slope is not a common characteristic in all the parts of the arch of a group. For instance, in the bottom plot we can see that \emph{Acturus} is very steep at $V_R = 30-40$\,\kms\, (orange lines) and flattens for the negative part of the arch ($V_R = -20$\,\kms, blue lines). We observe that secondary peaks have very different azimuthal and radial velocities when they start to show up at inner radius, but end up having very similar azimuthal velocity at larger radius. This corresponds to a flattening of the arches in the velocity distribution as $R$ increases. This is not observed in the simulations (Sect. \ref{sect:simulation}), and could be an effect of the centroid of the distribution dominating in case of undersampling in these regions.

As for the global shape of the lines, the resonant effects of the bar and spiral arms are expected to create kinematic substructures that, from epicyclic approximation analysis of the first order effects, have an almost constant vertical angular momentum $L_Z = R V_\phi$ \citep{sellwood2010lindblad,quillen2018coma}. Thus, if the moving groups have a bar resonance origin, their azimuthal velocities would naively be expected to follow lines $\propto R^{-1}$ (dashed grey lines in Fig.~\ref{fig:r_vphi_groups} top panel). In \citet{Ramos_2018} (their Fig.~6), this trend is observed for \emph{Hercules} and \emph{Hyades}. When extending the analysis to a larger region, we find that the groups deviate from the lines of constant angular momentum. 
In Appendix~\ref{sect:chisquared}, we compute the reduced chi-squared ($\chi^2_\nu$) parameter for all the groups. The only group that is statistically well approximated globally by this $V_\phi \propto R^{-1}$ trend is \emph{Arch/Hat}. We come back to this in Sect.~\ref{sect:discussion}.

The ridges are usually studied in $R$-$V_\phi$ diagrams colored by density or mean $V_R$ \citep[see Fig.~1 in][]{fragkoudi2019ridges}. By doing these projections, the complexity of the moving groups (arch curvature, bi-modalities, arch disruption, etc) is lost, offering only a partial understanding of the sample. With our methodology we can now visualize this complexity in a single plot. For instance, we can observe how the $V_R$-$V_\phi$ arch corresponding to the \emph{Acturus} moving group is a horizontal arch at $R=8$\,kpc (structures with different $V_R$ and common $V_\phi$) but it curves as we move to inner radius (the different structures fan out). This spreading clearly depends on the $V_R$, which is a sign of a curved arch. Mixed with \emph{Acturus}, we are able to observe the morphology of \emph{Bobylev} at $V_R>50$\,\kms. In the previously mentioned projections, the visualization of both structures is not possible, since the mean $V_R$ blends both contributions.

Beyond the detection and characterization of ridges in the radial direction, the main contribution of our method is the blind search of these kinematic structures in the three spatial dimensions at the same time. Next we focus on the representative of each moving groups and study their kinematics in the 3D space, azimuth submanifold ($Z=0$\,kpc) in Sect. \ref{sect:r_phi}, and vertical submanifold ($\phi=0$\,deg) in Sect. \ref{sect:r_z}.

%
%

\subsection{Azimuth submanifold} \label{sect:r_phi}

\begin{figure}
\centering
\includegraphics[width=0.49\textwidth]{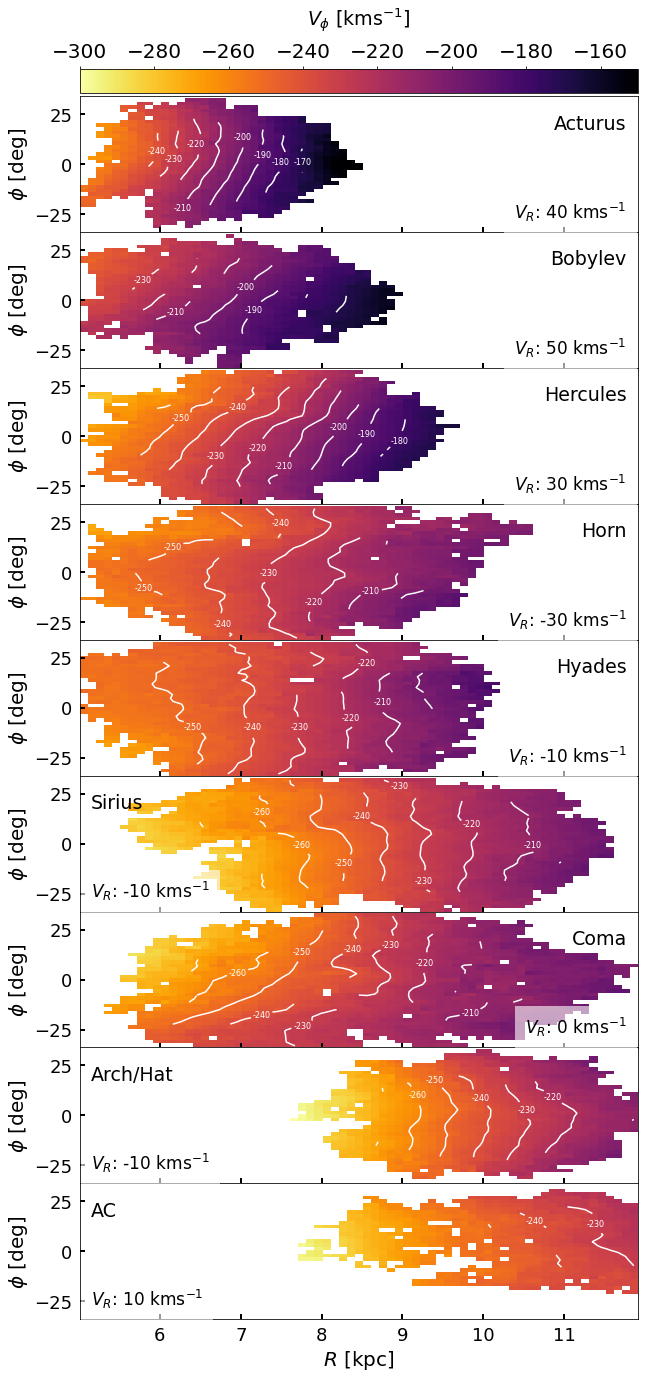}
\caption{Mean azimuthal velocity of the representative groups in the $R-\phi$ projection, for $|Z|<0.2$\,kpc. In white, the contours of regions with the same velocity are shown for clarity. \emph{Acturus}, \emph{Bobylev}, and \emph{Hercules} moving groups present a constant slope in the variation of azimuthal velocity along azimuth, whereas \emph{Horn}, \emph{Sirius} and \emph{Arch/Hat} moving groups present an axisymmetrical behaviour of the azimuthal velocity along azimuth.}
\label{fig:r_phi_vphi}
\end{figure}

We first do a cut in the structures around $Z=0$\,kpc ($|Z|<0.2$\,kpc) to observe the behaviour of the different moving groups as a function of $R$ and $\phi$. We obtain surfaces covering up to $\pm 25$ deg ($\approx 3.5$ kpc at Solar Radius). This is the first time that the moving groups are traced along the $Z=0$\,kpc plane with this completeness.

In Figure~\ref{fig:r_phi_vphi}, apart from the already studied decrease of $V_\phi$ with $R$, we can see how \emph{Acturus}, \emph{Bobylev}, and \emph{Hercules} moving groups present a slope in the azimuthal velocity along azimuth, whereas \emph{Horn}, \emph{Sirius} and \emph{Arch/Hat} moving groups present an axisymmetrical behaviour of the azimuthal velocity along azimuth, as expected in an axisymmetric potential.

It is interesting to quantify the variation of $V_\phi$ with $\phi$ for the structures. We evaluate this slope\footnote{Note that in our reference system a negative $\partial V_\phi / \partial \phi$ slope corresponds to an increase of $|V_\phi|$ with $\phi$ (a moving group with negative $\partial V_\phi / \partial \phi$ moves upwards in the velocity distribution with $\phi$).} $\partial V_\phi / \partial \phi$ at a given $R$ by restricting the structure to this $R$ value and doing a linear fitting of the surfaces in $\phi-V_\phi$. We compute the slope in the radius that minimizes the error in the fitting. The resulting values are: $-0.40$\,\kmsdeg\, at $R = 7$\,kpc for \emph{Acturus}, $-0.63$\,\kmsdeg\, at $R = 7$\,kpc for \emph{Bobylev}, $-0.50$\,\kmsdeg\, at $R = 8$\,kpc for \emph{Hercules}, $-0.04$\,\kmsdeg\, at $R = 10$\,kpc for \emph{Sirius}, and $-0.01$\,\kmsdeg\, at $R = 10$\,kpc for \emph{Arch}.

In \citet{monari2019hercules} they study the mean angular momentum evolution in $\phi$ for \emph{Hercules} in an analytical model. They predict that, in case of a co-rotation origin, the angular momentum $J_\phi$ of \emph{Hercules} at the solar radius must significantly decrease with increasing azimuth. Their model predicts the slope to be around $-8$\,kms$^{-1}$kpc\,deg$^{-1}$, and they observe a similar trend in Gaia DR2 data. Our equivalent value in angular momentum would be $-4$\,kms$^{-1}$kpc\,deg$^{-1}$, which is smaller than the predicted value.

In some parts of the disc, the mean azimuthal velocity in the plane for the total 6D sample decreases with increasing azimuth at a constant radius \citep[see Fig.~10 in][$R= 8-10$\,kpc]{katz2018dr2VR}. This is the behaviour that we observe for \emph{Acturus}, \emph{Bobylev}, and \emph{Hercules}. It would be worth investigating the relative contribution of each moving group to the total sample to understand the relation between these individual groups and the total average motion but we refer this to a future study.

%
%

\subsection{Vertical submanifold}\label{sect:r_z}

\begin{figure}
\centering
\includegraphics[width=0.48\textwidth]{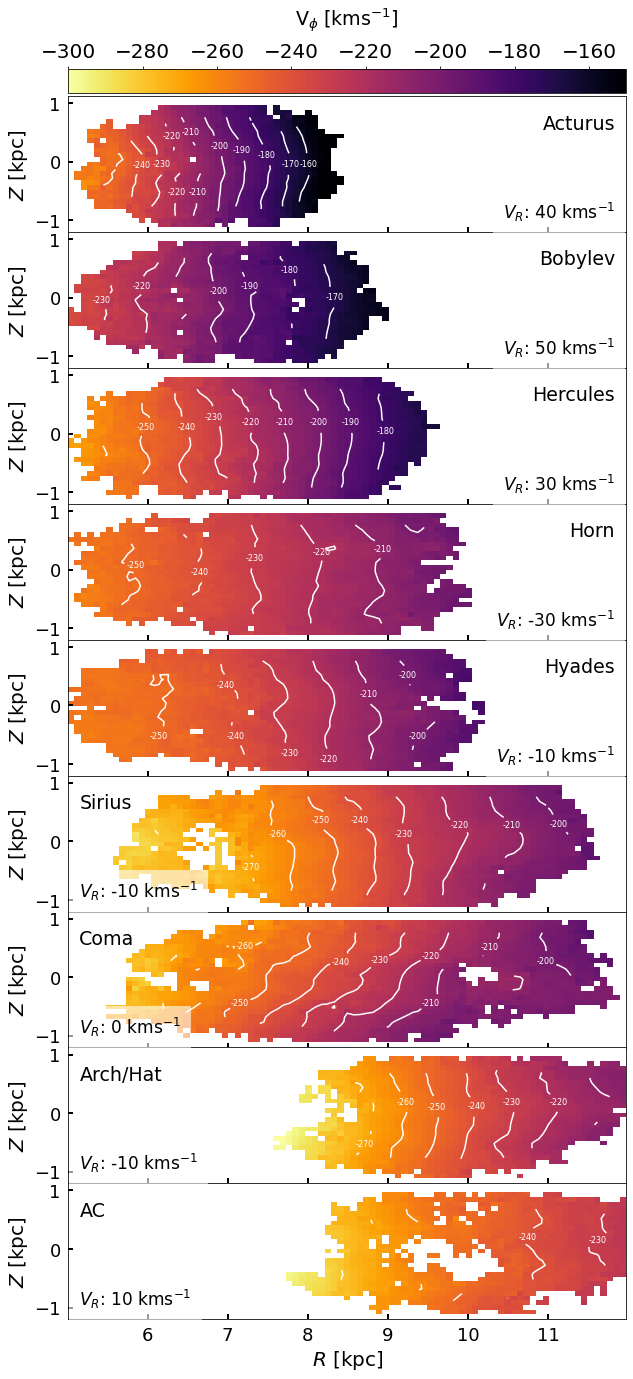}
\caption{Mean azimuthal velocity of the groups in the $R-Z$ projection, for $|\phi|<10$\,deg. In white, the contours of regions with the same velocity are shown for clarity. \emph{Coma Berenices} clearly presents an increasing $|V_\phi|$ with $Z$, and thus, strong vertical asymmetry. We measure a constant vertical slope of $\partial V_\phi / \partial Z = -15$\,\kmskpc. The rest of structures show vertical symmetry.}
\label{fig:r_z_vphi}
\end{figure}

We also study the projection in the $R-Z$ plane ($|\phi|<10$\,deg, see Fig.~\ref{fig:r_z_vphi}). In all the structures but \emph{Coma Berenices} (see below), we observe decreasing $|V_\phi|$ values for increasing $|Z|$. In addition, \emph{Acturus}, \emph{Bobylev}, and \emph{Hercules} --the same structures that show steeper slope in azimuth-- present a clear symmetry around $Z = 0$\,kpc. \emph{Arch} is also very simmetric in $Z$. In opposition, \emph{Horn}, \emph{Hyades}, and \emph{Sirius} present an steeper decrease in $|V_\phi|$ for $Z>0$\,kpc with respect to the other moving groups, and a more constant value for $Z<0$\,kpc. Finally, \emph{AC} has not enough signal at this point to analyse it properly.

\emph{Coma Berenices} clearly presents an increasing $|V_\phi|$ with $Z$, and thus, strong vertical asymmetry. We note that outside the range of $R=[7,10]$\,kpc, this moving group shows a change of behaviour in all the spatial projections (Figs. \ref{fig:r_vphi_groups}, \ref{fig:r_phi_vphi}, \ref{fig:r_z_vphi}) possibly because our methodology is linking it to other close structures. Therefore, focusing only on the $[7,10]$\,kpc, range, we measure a constant vertical slope of $\partial V_\phi / \partial Z = -15$\,\kmskpc, clearly different to the other groups and predictions from models with vertical symmetry.

It would be interesting to obtain a measurement of the vertical curvature of the moving groups at each radius, as done in the previous section with the slope in azimuth and with the vertical slope in \emph{Coma Berenices}. With noisy data, each order of derivatives increases its uncertainty and the measurements we obtained were not significant enough. In the future, with better data and/or a robust analytical model to fit the curvature at all radius in the same time, this measurement could be produced.

\section{Simulations}\label{sect:simulation}

\begin{figure*}
\centering
\includegraphics[width=0.77\textwidth]{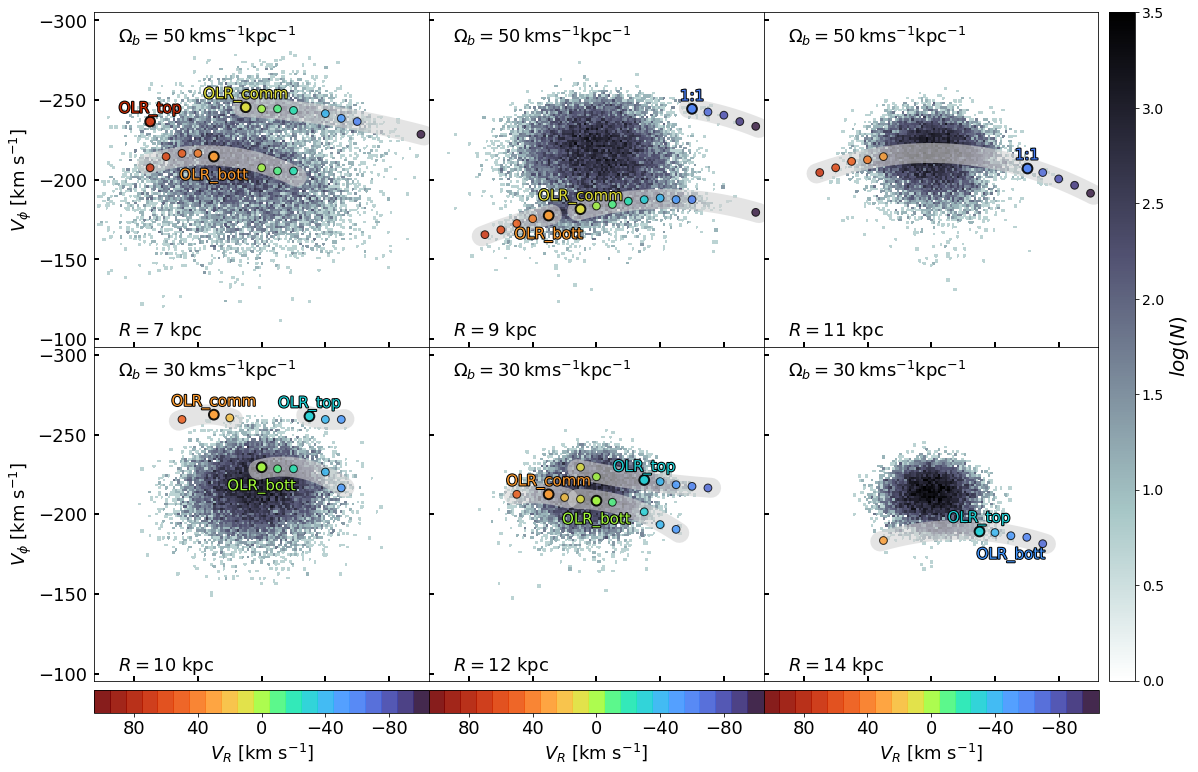}
\caption{Moving group detection in different neighbourhoods along the radial direction in the simulations, analog to Fig.~\ref{fig:SN_WT}. Top row: fast bar model. Bottom row: slow bar model.}
\label{fig:sim_WT}
\end{figure*}

\begin{figure*}
\centering
\includegraphics[width=0.82\textwidth]{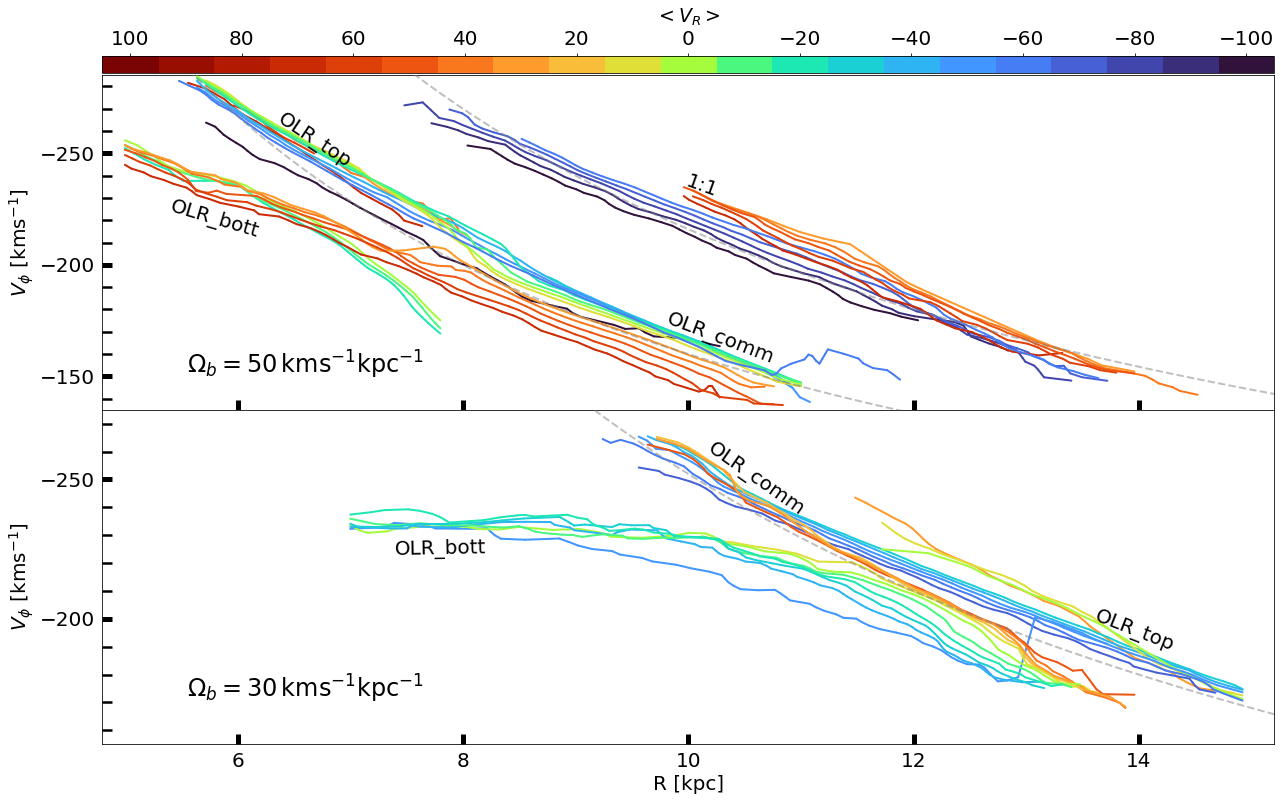}
\caption{Azimuthal velocity of the kinematic substructures in the radial direction ($\phi=0$º, $Z=0$\,kpc) for the test particle simulations, as a function of the radius, and coloured by their radial velocity. We include dashed grey lines corresponding to constant angular momentum lines as a guide. Top: Structures for the fast bar model. Bottom: Structures for the slow bar model. In the fast bar model, we are able to detect substructures related to the OLR and the $1:1$ resonance, in the slow bar we only detect structures related with the OLR. See also Fig.~\ref{fig:r_vphi_sim_meanVR} for the mean radial velocity histogram. With our methodology, we are able to show the complex morphology of the arches in a single image.}
\label{fig:r_vphi_sim}
\end{figure*}

In this section we apply the same methodology to simulations. This has two main goals: evaluating the performance of our method in a case where there are no selection effects, and allowing a comparison of the data to a model where a particular and known perturbation is present. In our case, we used a series of test particle simulations with 60M particles. The initial conditions and the Galactic potential are described in \citet{RomeroGomez2015}. In particular, the disc has a local radial velocity dispersion of $\sigma_{V_R}=30.3$\,\kms\, at the radius of $8.5$\,kpc. We integrate the initial conditions, first, in the axisymemtric potential of \citet{Allen1991} for $10\,$Gyr, and, secondly, we introduce the Galactic bar potential adiabatically during $4$ bar rotations ($2.46$\,Gyr for the slow bar and $1.47$\,Gyr for the fast bar), to finally integrate another $4$ bar rotations. The Galactic bar consists of the superposition of two aligned Ferrers ellipsoids \citep{Ferrers1877}, one modelling the triaxial bulge with semi-major axis of $3.13\,$kpc and the second modelling the long thin bar with semi-major axis of $4.5\,$kpc. We use two simulations, where the bar rotates as a rigid body with a constant pattern speed of $30$ and $50$\,kms\,kpc$^{-1}$. For the slow bar, the CR is located at $7.3$\,kpc and the OLR at $12.2$\,kpc. For the fast bar, the CR is located at $4.3$\,kpc and the OLR at $7.6$\,kpc. We use the final snapshot of the simulations, and assume that the bar is $30$\,deg in azimuth with respect to the Sun in the direction of rotation, close to the estimations for the MW \citep[][references therein]{blandhawthorn2016context}.

In these final snapshots, we execute the methodology described in Section \ref{sect:method} and obtain an optimal detection of the moving groups in a large range of the sample. These robust results, matching the predictions from previous works, validate the performance of our methodology in a known dataset. However, we find some substructures related to the centroid of the velocity distribution whose changes with azimuth, radius and height are mostly related to the rotation curve of the model. In this section we only show the substructure related with bar resonances and ignore the rest of groups extracted by the methodology. 

The selected moving groups are shown in the $V_R$-$V_\phi$ projection in Fig.~\ref{fig:sim_WT} (same as done in Fig.~\ref{fig:SN_WT} for the real data). In the simulation, the moving groups also show arches in this projection, which we are able to detect at several radii. Again, we can show the groups projected in the radial direction, (Fig.~\ref{fig:r_vphi_sim}, analog to Fig.~\ref{fig:r_vphi_groups}). In Figs.~\ref{fig:sim_WT} and \ref{fig:r_vphi_sim}, the top panels show the structures of the fast bar model (depicting the effects of the \emph{OLR} and \emph{1:1} resonance). The bottom panels of the figures show the structures of the slow bar model (only the effects of the \emph{OLR} appear). In the following sections, we analyse the fast bar model (Sect. \ref{sect:fast_bar}) and the slow bar model (Sect. \ref{sect:slow_bar}) in detail.

As explained in the introduction, a fast bar model places \emph{Hercules} near the OLR of the bar. In this model, \emph{Arch/Hat} could be explained as the 1:1 resonance. Instead, the slow bar model places \emph{Hercules} in the CR of the bar and \emph{Arch/Hat} in the OLR. Therefore, in this article we refer as Hercules-like to structures in the simulation which can be related to \emph{Hercules} in the data (i.e. generated by OLR for the fast bar model and by CR for the slow bar model), and Arch/Hat-like as the structures which can be related to the \emph{Arch/hat} (i.e. induced by the 1:1 for the fast bar model and by the OLR for the slow bar model).

In Fig.~\ref{fig:r_vphi_sim}, it is clear that the global position of the overdensities is not following exactly the lines of constant angular momentum predicted for small epicyclic amplitudes ($V_\phi\propto R^{-1}$). In the radial direction, the curvature of the \emph{OLR\_bott} structures is opposite in the two models. In addition, the fact that the structures merge at some radius is a clear evidence that they follow a trend different from $\propto R^{-1}$. The first order prediction for resonances is suboptimal for $V_\phi$ values far from the circular velocity. In Fig.~\ref{fig:r_vphi_sim}, we observe how the radial slope of the structures differs more from the prediction at small and large $V_\phi$ values.

\begin{figure}
\centering
\includegraphics[width=0.5\textwidth]{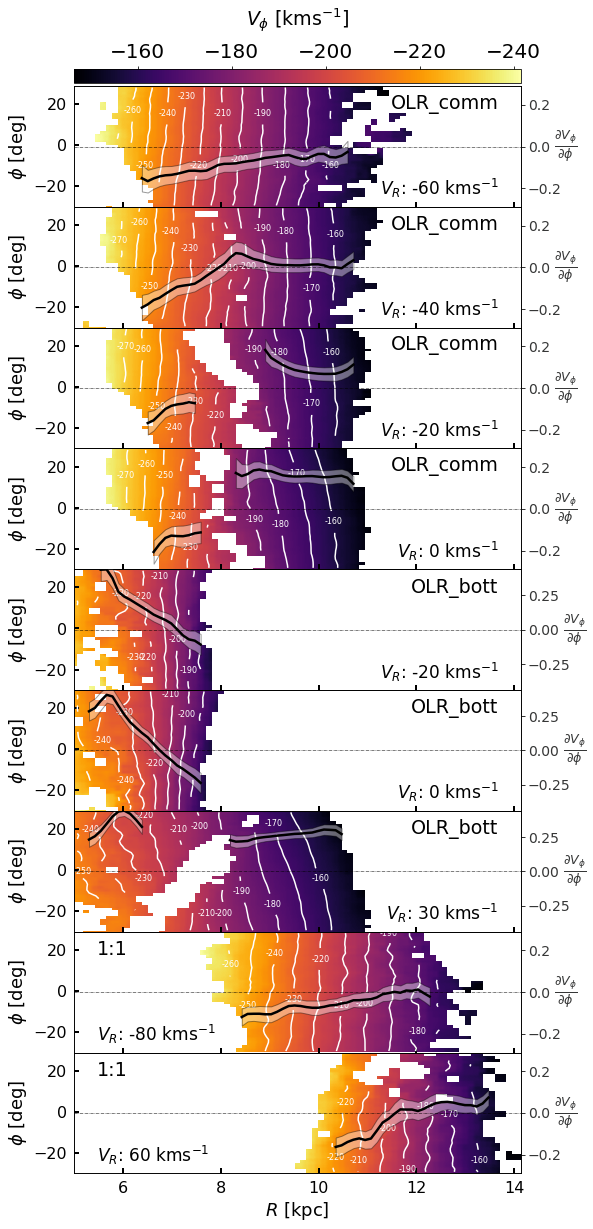}
\caption{Fast bar model. Mean azimuthal velocity of a selection of groups in the $R-\phi$ projection, for $|Z|<0.5$ kpc. In white, the contours of regions with the same velocity are shown for clarity. In black, the slope of the linear fitting $(\phi, V_\phi)$ is shown for every $R$ column, in units of $[$kms$^{-1}$deg$^{-1}]$. The $3\sigma$ error of the slope is shown in a translucent region around the line.}
\label{fig:r_phi_sim50}
\end{figure}

\subsection{Fast bar model}\label{sect:fast_bar}

In the fast rotating bar simulation ($\Omegab=50$\,\kmskpc), we are able to detect substructures related to the OLR and the 1:1 resonance with our methodology. As expected, both structures present an arch shape in the $V_R$-$V_\phi$ diagram (Fig.~\ref{fig:sim_WT}, top). The complete sampling and the simplicity of the simulation allows us to trace these arches unequivocally and observe their morphology in ranges up to $6$\,kpc in the radial and azimuthal directions, characterizing with strong robustness not only their position in the velocity space but the trend of the central and outer part of the arch separately. Note that the simulation has been integrated for 4 bar rotations, which places this model in the regime of a {\it young} fast bar, as defined in \citet{monari2019hercules}.

In the top part of Fig.~\ref{fig:r_vphi_sim}, the bi-modality 
of the OLR and the 1:1 resonance are clearly observed.
For the OLR, at $R<7$\,kpc we observe an Hercules-like arch (\emph{OLR\_bott}), extending to a maximum $V_R$ of $40$\,\kms\, (orange line on top). In this same region, we also observe a symmetric arch on top of the distribution tagged as \emph{OLR\_top}, with a maximum in $V_R=0$\,\kms\, (green line on top). At $R$ between $7$ and $8$\,kpc, the $V_R$-negative part of the bottom arch ($V_R$ around $-10$\,\kms) continues to decrease in $|V_\phi|$ and the positive part of the arch ($V_R>20$\,\kms) merges with the top arch (\emph{OLT\_top}) to form a unique structure covering the whole radial velocity range. This unique arch, \emph{OLR\_comm}, has its maximum at $V_R=-40$\,\kms\, (blue line on top). This arch configurations at $R = 7, 9,$ and $11$\,kpc can also be seen in the top row of Fig.~\ref{fig:sim_WT}.

The signature of the \emph{1:1} resonance (Arch/Hat-like structure) is less prominent than in the case of the \emph{OLR}. In Fig.~\ref{fig:sim_WT} (top row, $R=11$\,kpc), we observe a bi-modality in the histogram, with the upper part of the distribution being more prominent at negative $V_R$ values and the lower part more prominent in the positive $V_R$ range. In the $R$-$V_\phi$ diagram (Fig.~\ref{fig:r_vphi_sim}, top panel) we see that at inner radius we mostly detect negative $V_R$ and, at the outer parts, mostly positive $V_R$, giving more evidences of this bi-modality. Due to the small prominence of the resonance, we are not able to detect it when it is located in the centre of the distribution.

As we did with the real data, we exploit the 3D extent of the structures and show their trends in the azimuth submanifold. In Fig.~\ref{fig:r_phi_sim50} we show the $R-\phi$ projection of the azimuthal velocity for different parts of the arches (analog to Fig.~\ref{fig:r_phi_vphi}). The black lines in these panels show the azimuthal slope of $V_\phi$ in each radius (right vertical axes).

The \emph{OLR\_comm} structures present a discontinuity around $R=8$\,kpc. In Fig.~\ref{fig:r_vphi_sim}, this can be seen as the sudden drop in $V_\phi$ in the green-turquoise lines at $R\sim8$\,kpc ($V_R={-20,-10,0}$\,km\,s$^{-1}$). A similar discontinuity is present in two panels of the \emph{OLR\_bott} structure in Fig.~\ref{fig:r_phi_sim50}. The peak detection algorithm guarantees a minimum distance between peaks for robustness. Therefore, when two arches merge, it stops detecting one of them a while before the merge. Even with this, a few spurious detection can act as a bridge for the algorithm and and make it join both structures. Since the arches are merging, the algorithm described in Sect.~\ref{sect:BFS} (Alg.~\ref{algo:bfs_simple_interpolator}) detects a continuity and the structures are detected together.


When looking at Fig.~\ref{fig:r_phi_sim50}, it is important to remember that we are studying a {\it young} fast bar, which is known to introduce azimuth correlations, specially around the OLR \citep[see Fig.~A1 in][]{Trick_2021}.
For most of the structures, we see an azimuth slope of $V_\phi$ that depends on radius. Interestingly, some structures have negative $\partial V_\phi / \partial\phi$, such as \emph{Hercules} in the real data, and others have a positive slope. We observe three main structures with different patterns:

\begin{itemize}

     \item Rows 1-4, and 7. Upper part of the OLR bimodality. Before crossing the rotation curve ($R\approx 8$\,kpc), the $\partial V_\phi / \partial\phi$ slope of the structure is constant along all the $V_R$ values. For $R>8$\,kpc, a correlation appears between $\partial V_\phi / \partial\phi$ and $V_R$. This is a sign that the \emph{OLR\_comm} arch shown in Fig.~\ref{fig:sim_WT} (top row, $R=9$\,kpc) is moving to negative $V_R$ values in the $V_R$-$V_\phi$ diagram as $\phi$ increases (when $V_R$ is more positive, its $|V_\phi|$ decreases faster with $\phi$).
     
    \item Rows 5, 6, and 7 ($R<8$\,kpc). Hercules-like part of the distribution. The slope in azimuth decreases as $R$ increases. The contour lines merge when approaching the bar's long axis ($\phi_b = -30$\,deg).
    
    \item Rows 8 and 9. Arch/Hat-like part of the distribution. The slope in azimuth is constant along the different detected parts of the arch, and tends to flatten for increasing $R$ values.
\end{itemize}

\subsection{Slow bar model}\label{sect:slow_bar}

\begin{figure}
\centering
\includegraphics[width=0.5\textwidth]{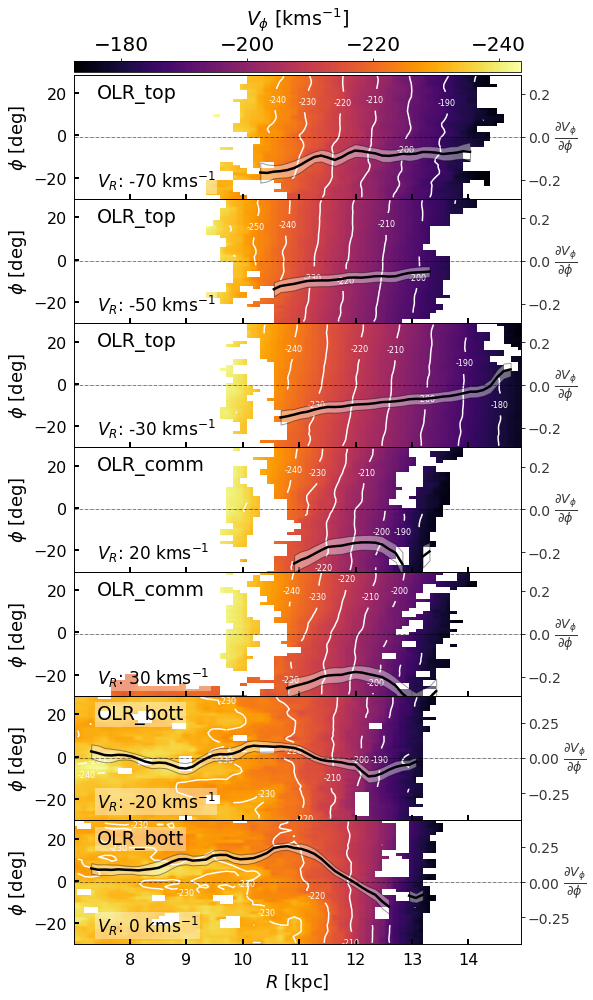}
\caption{Slow bar model. Mean azimuthal velocity of a selection of groups in the $R-\phi$ projection, for $|Z|<0.5$ kpc, analog to Fig.~\ref{fig:r_phi_sim50}. In this simulation, the shape of the arches is maintained along $\phi$ (same slope for all $V_R$ values), with a constant displacement to bigger $|V_\phi|$ as $\phi$ increases.}
\label{fig:r_phi_sim30}
\end{figure}

In the slow rotating bar simulation ($\Omegab=30$\,\kmskpc), we are able to detect the Arch/Hat-like overdensity caused by the OLR along a large range of radius, covering up to $6$\,kpc (Figs.~\ref{fig:sim_WT} and ~\ref{fig:r_vphi_sim}, bottom panels).
In the radial direction, we observe three different structures related with the OLR. Two of them have negative $V_R$ and the other in the positive $V_R$. The negative $V_R$ structures (\emph{OLR\_top} and \emph{OLR\_bott}) present a clear bi-modality in $V_\phi$. In the $V_R$-$V_\phi$ diagram (Fig.~\ref{fig:sim_WT}, bottom), the \emph{OLR\_top} forms a flat arch and the \emph{OLR\_bott} forms a curved arch. In the positive $V_R$ regime (orange-red lines), we observe one single structure (\emph{OLR\_comm}) compatible with \emph{OLR\_top} at $R=10$ to $11$\,kpc which continues to decrease its azimuthal velocity with $R$ in a constant slope and merges \emph{OLR\_bott} from $R=12$\,kpc to the outer parts of the disc.

Again, we can exploit the 3D extension of the manifolds to study how these groups are evolving in the $R-\phi$ projection. In Fig.~\ref{fig:r_phi_sim30} (analog to Fig.~\ref{fig:r_phi_sim50}, but for the slow bar model) we observe two main structures, corresponding to the different parts of the OLR bi-modality (whose upper part corresponds now to the Arch-like group):
\begin{itemize}
    \item Rows 1-3. Top part of the OLR bi-modality. It presents a constant slope in azimuth ($\partial V_\phi / \partial \phi = -0.1$\,kms$^{-1}$deg$^{-1}$ at $R = 12$\,kpc), flattening as $R$ increases, similar to the Arch/Hat-like observed in Fig.~\ref{fig:r_phi_sim50} for the fast bar.
    \item Rows 4-7. Bottom part of the bi-modality. The slope is less constant, and we observe the same decrease in $R$ as in the Hercules-like part of the fast bar model. 
\end{itemize}

We do not detect overdensities caused by CR. In general, the expected Hercules-like moving group caused by CR is less prominent than when formed by the OLR \citep{binney2018orbits, hunt2018hercules}, especially if higher order modes are missing from the bar model. Also, if the velocity dispersion of the disc is too large in this region we could have less trapping and less resolution. Another explanation could be that the length of the bar in our models does not favour CR trapping. In the future, we aim to perform the same analysis with other models to analyse this effect \citep[e.g.][]{sormani}. 

In Fig.~\ref{fig:r_vphi_sim_meanVR}, we show the $V_\phi$-$R$ projection colored by mean $V_R$. In that figure we observe a very faint red overdensity in the CR region, but much weaker compared to the rest of the detected resonances in the fast and the slow bar models.

\section{Discussion} \label{sect:discussion}

Simulations which contain only a bar perturber offer us the possibility to characterize its effects in a robust way.
With it, we can then compare the manifolds extracted from the simulation with the ones seen in the data. As explained in the introduction, one of our goals with the characterization of the manifolds is to search for kinematic observables that distinguish resonances from short/fast or long/slow bars.

\paragraph{Radial direction.}

First, we focus on the slope, shape, and evolution of the moving groups in the data and the simulation along the radial direction (Figs. \ref{fig:r_vphi_groups} and \ref{fig:r_vphi_sim}).

The global gradient of the structures in the radial direction deviates from the naive first order prediction for resonances (curves following $V_\phi\propto R^{-1}$) in the data and the simulations, as expected in more realistic cases, even if they have a resonance origin.
In the simulations, we see that the lines are less curved than the simple prediction and some parts of the \emph{OLR\_bott} have even the opposite curvature. This is also observed in the data, where all the moving groups are less steep in the inner parts of the disc than the prediction, and \emph{Hercules} and \emph{Acturus} present the same opposite curvature pattern as the \emph{OLR\_bott}.

We see that, in the $V_\phi$-$R$ projection colored by mean $V_R$ of the slow bar scenario of our simulations (Fig.~\ref{fig:r_vphi_sim_meanVR}) the co-rotation overdensity is not significant enough. This has been already discussed in the introduction and in the previous section, and could be related to the specificities of the model considered here.
In the fast bar scenario an Hercules-like structure is formed at $R=[6,8]$\,kpc, which shows an arch shape with a maximum in $V_R = 40$\,\kms\, (Fig.~\ref{fig:sim_WT}, top left panel), coherent with the \emph{Hercules} moving group in the data. In this same radial region the top part of the OLR forms an arch in negative $V_R$, which can be related to \emph{Horn}. Finally, from $R=8$\,kpc towards the outer parts of the disc, the OLR forms a single arch with a maximum in $V_R = -20$\,\kms. A similar arch, present at all the radial velocities, was already observed in the models by \citet{boby2010hercules}. In the data this region ($R>9$\,kpc, $V_\phi<-170$\,\kms) contains very few stars and we did not detect any group there.

The other feature commonly associated with bar resonances is the \emph{Arch/Hat} moving group. For a fast bar, it can be explained by the 1:1 resonance trapping region, and for a slow bar it matches the position of the OLR. In the data, the number of stars with $R>10$\,kpc is low. Therefore, the quality of the shape characterization of \emph{Arch/Hat} moving group is poor. Even so, in the panel $R=10.4$\,kpc of Fig.~\ref{fig:SN_WT} our method finds a tentative arch split around $V_R=10$\,\kms. This would match the split we detect between the \emph{OLR\_comm} and \emph{OLR\_top} structures in the slow bar model (Fig.~\ref{fig:r_vphi_sim}, bottom panel at $R=11$\,kpc). 
In the fast bar simulation, the prominence of the 1:1 resonance is low and we can only detect it in regions far from the centroid of the distribution. Even with this limited detection, we do not observe a bi-modality in the negative $V_R$ region, which would be a way to discriminate between resonances.

\paragraph{Azimuth submanifold.}

In both simulation models, for the Arch/Hat-like moving groups we observe a constant positive slope of $|V_\phi|$ in azimuth which tends to flatten (get closer to $0$\,\kmsdeg) as $R$ increases. In these Arch/Hat-like groups, the slope of the azimuthal velocity in azimuth does not depend on $V_R$ (Figs.~\ref{fig:r_phi_sim50} and \ref{fig:r_phi_sim30}). In the data (Fig.~\ref{fig:r_phi_vphi}), the \emph{Arch/Hat} group is very noisy.
Therefore, as discussed in the previous paragraph it is still complicated to use the \emph{Arch/Hat} velocities for a final relation to a specific resonance.

In Fig.~\ref{fig:r_phi_sim50}, we can observe the different parts of the OLR resonance in the fast bar. The Hercules-like overdensity ($R=[6,8]$\,kpc) shows a constant positive slope of $|V_\phi|$ in azimuth common along all $V_R$. In opposition, the trapping region of the resonance \emph{OLR\_comm} has a slope which depends on the $V_R$. This can be interpreted as the arch moving along $V_R$ in the $V_R$-$V_\phi$ diagram when moving in azimuth, which is observed at the OLR in other simulations of a young bar far from phase-mixed \citep{dehnen2000effect, boby2010hercules,Trick_2021}. Since we have the complete information of the moving groups in the data, we can reproduce the same analysis of the moving group slope at different $V_R$ for some moving groups in the data. In Figures \ref{fig:r_phi_vphi_Hercules} and \ref{fig:r_phi_vphi_Hyades}, we show this slope for \emph{Hercules} and \emph{Hyades}, respectively. We do not observe this dependence in $V_R$ for any of the groups. In the velocity distribution, this can be observed as the moving group arches increasing their $|V_\phi|$ along azimuth, but no displacement along $V_R$.

\paragraph{Vertical submanifold.}

For the resonces of the bar, the vertical displacement of $V_\phi$ should be dominated to first order by the vertical potential of the Galaxy \citep{alKazwini2022}. In our results, the vertical curvature of all the moving groups (except for \emph{Coma Berenices}) at a common radius is very similar, thus matching the analytical prediction. This means that this data is a good candidate to constrain the 3D structure of the potential. To second order, we could try to distinguish  different curvatures of different resonances. In \citet{alKazwini2022}, the displacement of the resonances at $Z=1$\,kpc with respect to the galactic plane is measured to be $8$\,\kmskpc\, for the corotation, $6$\,\kmskpc\, for the OLR, and $4$\,\kmskpc\, for the 1:1 resonance. Our maximum resolution, given by the $V_\phi$ histogram at each pixel, is $1$\,\kmskpc. Therefore, disentangling which resonances create each moving group with the vertical information is beyond our current capabilities.

In the vertical behaviour of the moving groups, a clear outlier is \emph{Coma Berenices}. In \citet{quillen2018coma} \citep[also][]{monari2018coma, laporte2019footprints}, the \emph{Coma Berenices} moving group is observed to be present only at negative galactic latitudes, showing evidence for incomplete vertical phase-mixing. With the new EDR3 data and our methodology, we are able to detect the group in a larger extension and at positive and negative galactic latitudes but it does show a clear vertical asymmetry in its azimuthal velocity. We measure a constant vertical slope of $\partial V_\phi / \partial Z = -15$\,\kmskpc. In the plots of a phase spiral colored by $V_\phi$ \citep{antoja2018dynamically}, it is seen that there must be a correlation between $Z$ and $V_\phi$ (higher $|V_\phi|$ values at positive $Z$). This slope in \emph{Coma Berenices} could be a projection of this correlation.

%
%

\section{Conclusions}\label{sect:conclusions}

We have sampled, with Gaia EDR3 6D data, the manifolds tracing the main moving groups in the solar neighbourhood along the $(R,\phi,Z,V_R,V_\phi)$ space, in an automatic way. We have revealed the skeleton of the velocity distribution in a multidimensional space that we can then explore along the radial direction, and characterize in the azimuth and vertical submanifolds.
This methodology has been successfully tested with two simulations of the effects of a (dynamically young) bar. We have been able to observe and quantify the spatial evolution of the observed moving groups in a large range of about $3$\,kpc around the sun. Our main results and conclusions are:
\begin{itemize}
    \item The azimuthal velocity of the moving groups in the radial direction does not follow lines of constant angular momentum, deviating from the naive first order prediction for resonances. In the simulations, resonant structures also deviate from this simple prediction, demanding more complex analytic predictions.
    \item The spatial evolution of the moving groups is complex.
    The moving groups configuration observed in the SN is not maintained throughout the disc. The relative position between the arches and their curvature changes across space, and the different moving groups split and merge several times. This is expected in a context of bifurcating orbital families, for example in the case of resonances.
    \item In our slow bar simulation, we observe a bi-modality created by the OLR in the outer parts of the disc. This bi-modality is also observed in the \emph{Arch/Hat} moving group in the data. This intriguing agreement could favour the slow bar scenario, and opens the possibility of a test with future data.
    \item The \emph{Acturus}, \emph{Bobylev}, and \emph{Hercules} moving groups present a positive slope of their $V_\phi$ location with the azimuth. We measure this slope to be $-0.50$\,\kmsdeg\, at $R = 8$\,kpc for \emph{Hercules}.
    \item The azimuthal velocity of the \emph{Horn}, \emph{Sirius} and \emph{Arch/Hat} moving groups presents an axisymmetrical behaviour. In both our simulations, we observed a small azimuthal gradient in $V_\phi$ in the Arch/Hat-like structures, although it approches $0$ as $R$ increases. This could be related to the young bar model we are using.
    \item The vertical curvature of the moving groups is similar at the same $R$. These curvatures are dominated by the gravitational potential to first order, independently of the observed resonance. However, we notice that the \emph{Coma Berenices} group deviates from this behaviour, which points to a different dynamical origin that deserves further investigation.
    \item In the fast bar simulation, a correlation between $\partial V_\phi / \partial \phi$ and $V_R$ is observed for the OLR trapping region.
    The region where this correlation is observed in the simulation ($R>9$\,kpc, $V_\phi<-170$\,\kms) is poorly sampled in Gaia EDR3, but this could potentially be used to give information on the pattern speed of the bar with better data.
\end{itemize}

Spiral arms, resonances with the bar, accretion events, and possibly other effects can contribute to the present phase space distribution from which we obtain our observables.
Disentangling all the contributions of these dynamical processes is hard to address. In this work, we have shown the complexity of the phase-space structure that even a single mechanism (namely the bar) can produce. Our methodology allows to extract a quantitative and robust measurement of the observed phase space substructure that can be then compared and/or fit to different models. 

In this paper, we have developed the methodology for the study of the disc with Gaia data, and its formulation is easily generalizable. The same approach can be exported to other substructures in astrophysics (e.g. blind search for streams and shells in the halo) and even datasets outside this field.

In the following months, Gaia DR3 will revolutionize once again our field of research. The new 6D sample contains about 33M stars, covering a significantly larger region of the disc. During the review process, this new data has been used in \citet{luccini2022movinggroups} to reveal four new moving groups candidates in the solar neighbourhood. With this work, we are ready to process the Gaia DR3 data and extract its full potential.


\begin{acknowledgements}
We thank the anonymous referee for her/his helpful comments. This work has made use of data from the European Space Agency (ESA) mission {\it Gaia} (\url{https://www.cosmos.esa.int/gaia}), processed by the {\it Gaia} Data Processing and Analysis Consortium (DPAC, \url{https://www.cosmos.esa.int/web/gaia/dpac/consortium}). Funding for the DPAC has been provided by national institutions, in particular the institutions participating in the {\it Gaia} Multilateral Agreement. This work was (partially) funded by the Spanish MICIN/AEI/10.13039/501100011033 and by "ERDF A way of making Europe" by the “European Union” through grant RTI2018-095076-B-C21, and the Institute of Cosmos Sciences University of Barcelona (ICCUB, Unidad de Excelencia ’Mar{\'\i}a de Maeztu’) through grant CEX2019-000918-M.
MB acknowledges funding from the University of Barcelona’s official doctoral program for the development of a R+D+i project under the PREDOCS-UB grant.
TA acknowledges the grant RYC2018-025968-I funded by MCIN/AEI/10.13039/501100011033 and by ``ESF Investing in your future''. BF, GM and PR acknowledge funding from the Agence Nationale de la Recherche (ANR project ANR-18-CE31-0006 and ANR-19-CE31- 0017) and from the European Research Council (ERC) under the European Union’s Horizon 2020 research and innovation programme (grant agreement No. 834148).
\end{acknowledgements}


\bibliographystyle{aa}
\bibliography{MyBib}


\begin{appendix}

\section{Distance Bias}\label{sect:distance_bias}

We should be more worried about systematic errors in distance than about random errors since those would in principle only blur the structures. Systematic errors in distance may arise from the conversion from parallax to distance, the use of priors in the Bayesian procedures, etc. While it is difficult to evaluate the systematics in the distances that we used, one can perform some simple tests. In order to estimate the impact of a systematic error in the distance estimation we regenerated the sample using two worst case scenarios: an overestimation of all distances by $+10\%$ and an underestimation by $-10\%$. When comparing distances from StarHorse \citep{starHorse2022} and the photogeometric distances from \citet{bailer2021estimating} we see that for $80\%$ of the stars, the distance differences are smaller than $4\%$ at $2$\,kpc and smaller than $7\%$ at $3$\,kpc. The error that we consider in this test is thus larger than this difference.

\begin{figure}
\centering
\includegraphics[width=0.48\textwidth]{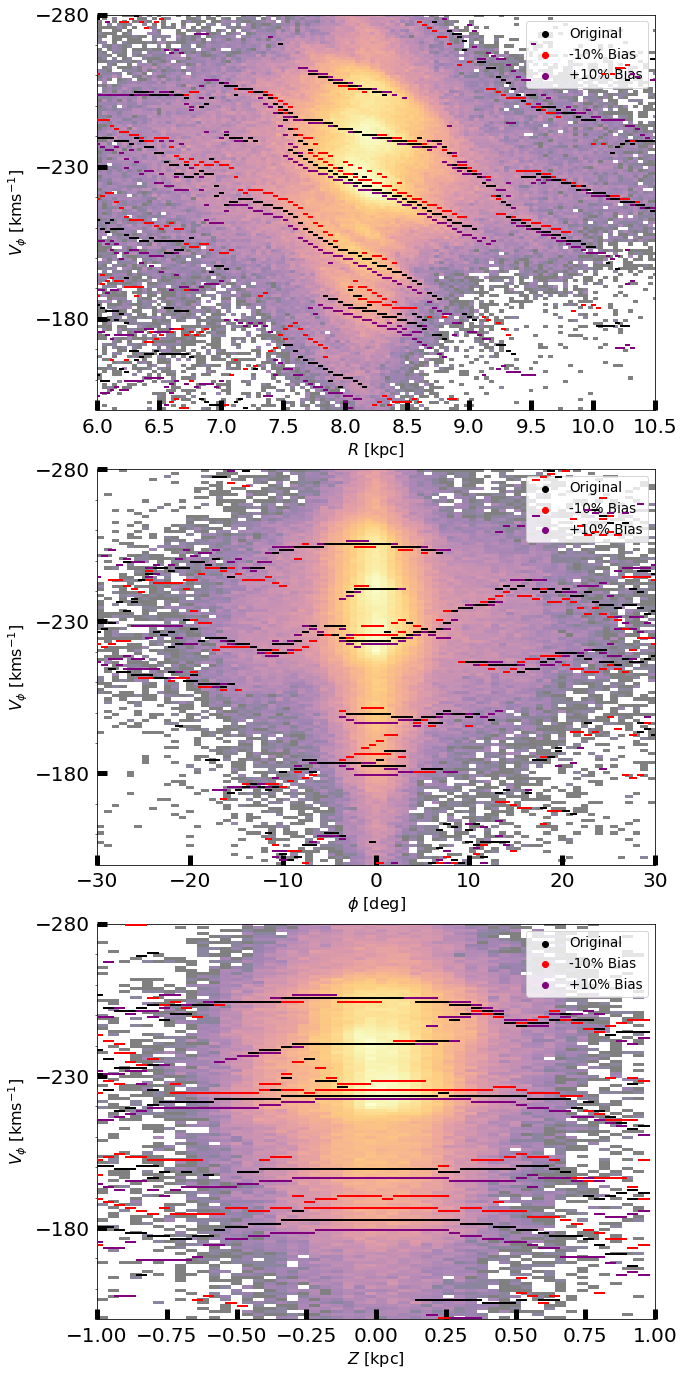}
\caption{Peak detection of the methodology for the Original, $-10\%$ bias, and $+10\%$ bias samples, in the Radial, Azimuthal, and Vertical directions.}
\label{fig:distance_bias}
\end{figure}

In Fig.~\ref{fig:distance_bias}, we show the results of the peak detection part of our methodology in the radial, azimuthal and vertical directions for the three samples (Original, $-10\%$ bias, and $+10\%$ bias). In it, we observe that introducing this bias produces a contraction or expansion of some dimensions of the phase-space, thus slightly modifying the position of the moving groups.

Quantifying the difference in the position of the peaks, we can estimate the error that this bias could introduce. In general, the bias is more significant if the azimutal velocity of the group is far from the solar azimuthal velocity. 
For ridges in the $|V_\phi|<|V_{\phi,\odot}|$ region, an underestimate of the distance increase the $|V_\phi|$ of the ridge, and a overestimate will decrease the ridge $|V_\phi|$. For Hercules, this produces a bias of $3$\,\kms\,in the SN ($V_\phi = -200$\,\kms), for Acturus, the bias increases up to $8$\,\kms.
The opposite effect is seen for stars in the $|V_\phi|>|V_{\phi,\odot}|$ region. For the Arch/Hat, this bias is $5$\,\kms\,at $R=9$\,kpc ($V_\phi=-280$\,\kms).
It is important to notice that we only detect kinematic structures at these low and high azimuthal velocities close to the Sun, precisely in the region where a bias in the distance measurement is less likely, mitigating the possibility of a distance bias.

In the radial direction, both $R$ and $V_\phi$ are proportional to the distance. Therefore, when computing the slope $\partial V_\phi / \partial R$ the dependence on distance cancels out. This is why the slopes that we observe for the structures in the plane $R$-$V_\phi$ (ridges) are maintained when the bias in the distance is introduced.
In the azimuthal direction, an underestimation/overestimation of the distance produces a small negative/positive curvature of $2$\,\kms\,at $|\phi| = 20$\,deg. In the vertical direction, an underestimate/overestimate of the distance produces a small positive/negative curvature of $2$\,\kms\,at $Z=\pm0.7$\,kpc.

\section{Epicyclic approximation discrepancy}\label{sect:chisquared}

\begin{table}[]
\setlength{\tabcolsep}{10pt}
\caption{Reduced Chi-Squared ($\chi^2_\nu$) parameter for each group, sorted in increasing order.}
\label{tab:chi_squared}
\centering
\begin{tabular}{ll}
\hline \hline
\rule{0pt}{12pt}\textbf{Group}    & $\chi^2_\nu$ \\[2pt] \hline
\textit{Arch/Hat} & $3.60$                \\
\textit{Acturus}  & $20.52$               \\
\textit{Bobylev}  & $31.12$               \\
\textit{AC}       & $32.44$               \\
\textit{Hercules} & $89.54$               \\
\textit{Coma}     & $148.61$              \\
\textit{Sirius}   & $216.46$              \\
\textit{Hyades}   & $252.12$              \\
\textit{Horn}     & $259.56$              \\ \hline
\end{tabular}
\end{table}

Here we test whether the structures in the radial direction (ridges) follow lines of constant angular momentum. To do this we use a reduced chi-squared ($\chi^2_\nu$) test, considering a constant error in the detection of $\sigma = 5$\,\kms, which is half of the largest scale used in the wavelet. We compute the $L_z$ which provides the best $V_\phi=L_z/R$ fitting for each group (i.e. minimizes the $\chi^2_\nu$ parameter). The best fitting curves are shown in Fig.~\ref{fig:r_vphi_groups}.
In Table~\ref{tab:chi_squared}, we show the $\chi^2_\nu$ parameter for each group. With it, we can now compare the goodness of fit for each structure. There are three groups of results. In the first one the Arch/Hat group is clearly well fitted by a line of constant angular momentum. A second group, formed by Acturus, Bobylev, and AC presents low chi-squared values but clearly incompatible with an optimal fitting. Finally, the shape of the rest of the structures is absolutely different from lines of constant angular momentum.

\section{Azimuth submanifold along a group}\label{sect:Hercules_Hyades_azimuth}

\begin{figure}
\centering
\includegraphics[width=0.48\textwidth]{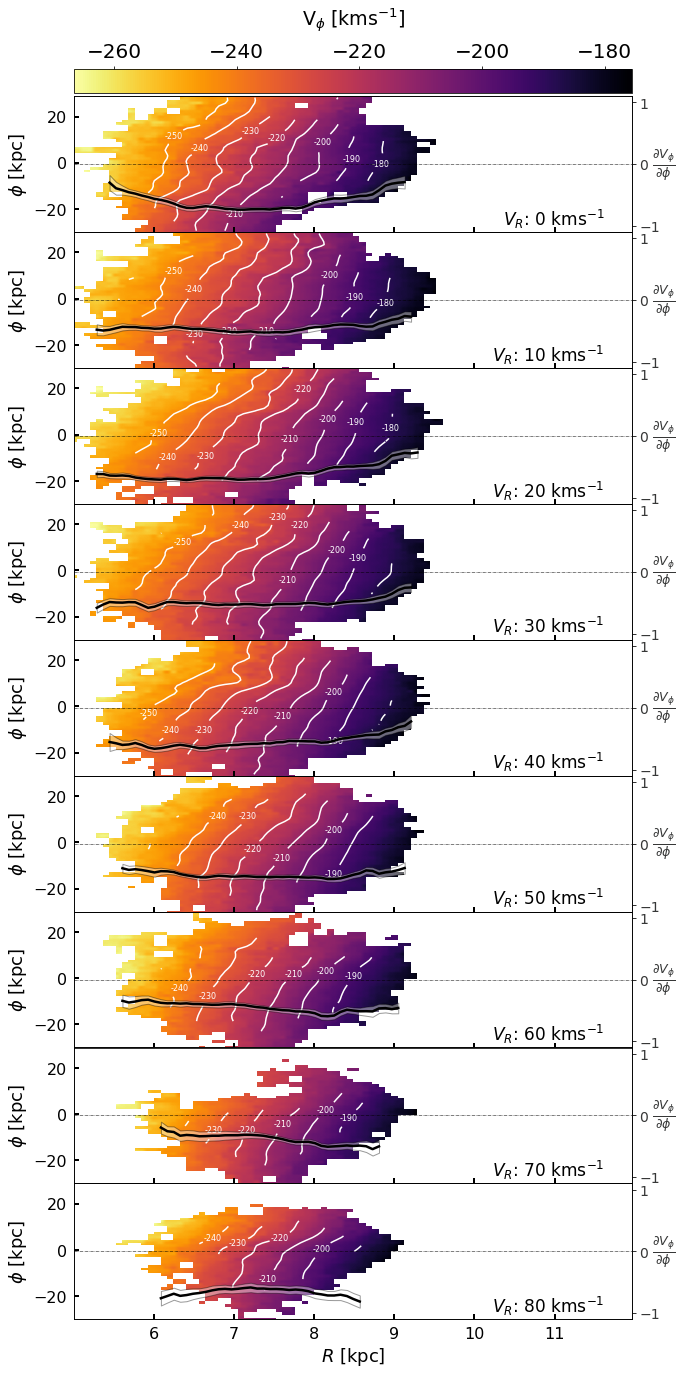}
\caption{Mean azimuthal velocity of the groups in \emph{Hercules} in the $R-\phi$ projection, for $|Z|<0.2$ kpc. In white, the contours of regions with the same velocity are shown for clarity. In black, the slope of the linear fitting $(\phi, V_\phi)$ is shown for every $R$ column, in units of $[$kms$^{-1}$deg$^{-1}]$. The $3\sigma$ error of the slope is shown in a translucent region around the line. The group has the same slope at all $V_R$. This corresponds to a vertical displacement of the moving group in the $V_R$-$V_\phi$ diagram.}
\label{fig:r_phi_vphi_Hercules}
\end{figure}

In Figure~\ref{fig:r_phi_vphi} we show the azimuth submanifold of a representative for each group. In this appendix, we extend this analysis and show how each part of \emph{Hercules} and \emph{Hyades} arches is evolving. In addition, since we have already shown these plots we include the extra information on the slope of $V_\phi$ in azimuth at each radius. With it, we can compare this groups to the resonances studied in the simulations (Figs. \ref{fig:r_phi_sim50} and \ref{fig:r_phi_sim30}).

The \emph{Hercules} moving group (Fig.~\ref{fig:r_phi_vphi_Hercules}) presents a stable and constant slope of $V_\phi$ in azimuth at the centre of the sample ($R=[6.5,8]$\,kpc). In the limits of the sample, where the significance of the group is smaller compared to the sample, this slope tends to flatten for the regions with $V_R\approx0$ and to become more steep in the large $V_R$ parts of the arch.

\begin{figure}
\centering
\includegraphics[width=0.48\textwidth]{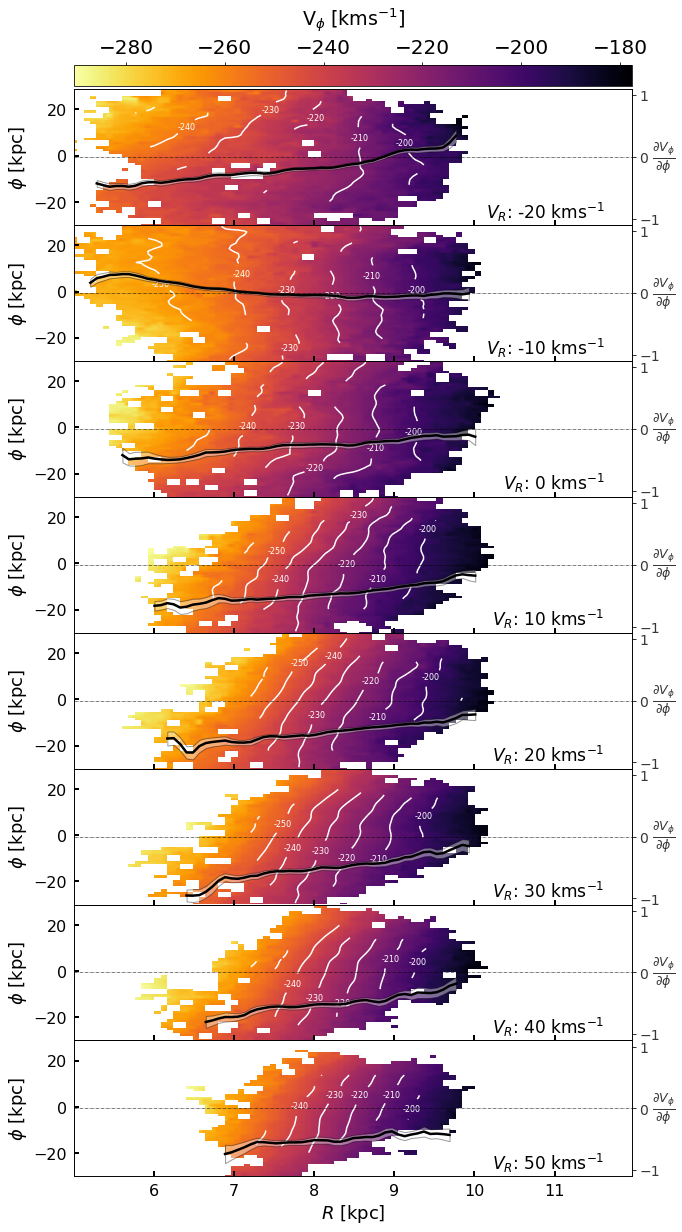}
\caption{Mean azimuthal velocity of the groups in \emph{Hyades} in the $R-\phi$ projection, for $|Z|<0.2$ kpc. Analogus to Fig.~\ref{fig:r_phi_vphi_Hercules}.}
\label{fig:r_phi_vphi_Hyades}
\end{figure}

The \emph{Hyades} moving group (Fig.~\ref{fig:r_phi_vphi_Hyades}) also presents a stable behaviour at all $V_R$. In the negative $V_R$ end of the arch (top two rows in the figure) the group has a very low significance, thus leading to a noisy detection. 


\section{Extra projections of the simulations}\label{sect:Extra_projections}

\begin{figure*}
\centering
\includegraphics[width=\textwidth]{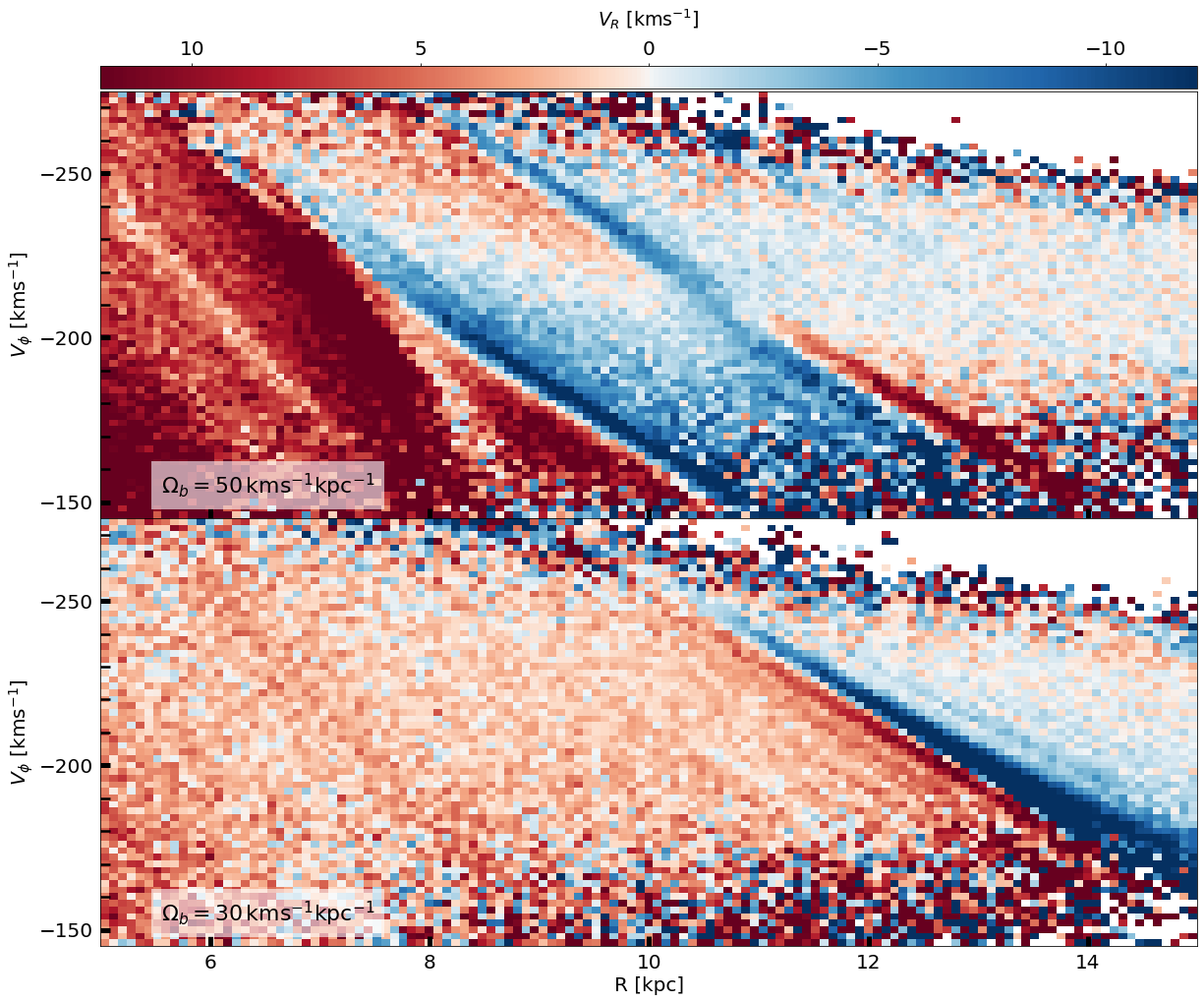}
\caption{Mean radial velocity of the kinematic substructures in the radial direction ($\phi=0$º, $Z=0$\,kpc) for the test particle simulations, as a function of the radius, and the azimuthal velocity. Top: Fast bar simulation. Bottom: Slow bar simulation.}
\label{fig:r_vphi_sim_meanVR}
\end{figure*}

It is the first time that a simulation is studied using the projection shown in Fig.~\ref{fig:r_vphi_sim}. In general, these studies are done in projections of $\langle V_R\rangle$. In order to compare both results, in Fig.~\ref{fig:r_vphi_sim_meanVR} we show this projection for the simulation.

In the top panel of Fig.~\ref{fig:r_vphi_sim_meanVR}, in red, we see the Hercules-like overdensity, which steeply decreases in $V_\phi$ around $R = 8$\,kpc. The upper part of the OLR bi-modality continues to decrease in a less steep trend, with negative $\langle V_R\rangle$ values on top and negatives in the bottom of the resonance. Finally, in the outer part of the disc we  observe the effect of the 1:1 resonance, which shows a swap in $\langle V_R\rangle$ sign when crossing the rotation curve. 

As for the bottom panel of Fig.~\ref{fig:r_vphi_sim_meanVR}, co-rotation should appear at $R=7.3$\,kpc, but we can not see any significant structure in this region. In the outer parts of the disc we do observe the OLR placed at $12.2$\,kpc. Bottom: Fast bar simulation. We observe a lot more substructure. CR ($R_{CR}=4.3$\,kpc) can be seen as a stripe at inner radius, although not significant enough to be detected by our methodology. At solar radius, we observe the effect of the OLR ($R_{OLR} = 7.6$\,kpc).

\end{appendix}

\end{document}